\documentclass[a4paper,11pt]{article}
\usepackage{amsbsy}
\usepackage[fleqn]{amsmath}
\usepackage{amssymb}

\usepackage{array}
\usepackage{booktabs}
\usepackage{xcolor}
\definecolor{linkred}{RGB}{128,0,32}
\usepackage[colorlinks=true,linkcolor=linkred,citecolor=linkred,urlcolor=linkred]{hyperref}

\usepackage{mathrsfs}
\usepackage{tabularx}
\usepackage{latexsym}
\usepackage{textcomp}
\usepackage{pstricks}
\usepackage{colortbl}
\usepackage{multirow}
\usepackage{fancyhdr}
\usepackage{bm}
\usepackage{booktabs}
\usepackage{subfigure}
\usepackage{caption}
\usepackage{soul}
\usepackage{cleveref}
\usepackage{cite}
\usepackage[pagewise]{lineno}
\usepackage{ulem}
\usepackage{graphicx}
\usepackage{epstopdf}
\usepackage{indentfirst}
\usepackage{setspace}
\usepackage[latin1]{inputenc}
\usepackage{titlesec}
\usepackage{tabularx}
\usepackage{ragged2e}
\usepackage{booktabs}
\usepackage{qcircuit}
\usepackage{braket}
\usepackage{svg}
\usepackage{amsmath}

\usepackage{algorithm}
\usepackage{algpseudocode}
\usepackage{mdframed}
\usepackage{xcolor}

\titleformat*{\section}{\large\bf}
\titleformat*{\subsection}{\normalsize\bf}
\crefname{section}{Section}{Sections}
\crefname{equation}{Eq.}{Eqs.}
\crefname{figure}{Figure}{Figs.}
\crefname{table}{Table}{Tables}
\crefname{algorithm}{Algorithm}{Algorithms}
\crefname{appendix}{Appendix}{Appendices}
\Crefname{appendix}{Appendix}{Appendices}
\usepackage{setspace}
\usepackage[absolute,overlay]{textpos}

\usepackage{fancyhdr}

\begin{document}

\thispagestyle{empty}

\begin{center}
\Large\textbf{Nonlinear path-following via the asymptotic numerical method
\\on a quantum processor}
\end{center}

\normalsize{
\begin{center}
\setcounter{footnote}{1}Yongchun Xu$^{\S,}$\footnote{These authors contributed equally to this work.\label{co}},
Zengtao Kuang$^{\S,}$$\textsuperscript{\ref{co}}$, 
Qun Huang$^{\S}$, 
Jie Yang$^{\S}$,
Hamid Zahrouni$^{\#}$, \\
Michel Potier-Ferry$^{\#}$,
Kaixuan Huang$^{\lozenge}$, 
Jia-Chi Zhang$^{\ddagger}$, \\
Heng Fan$^{\ddagger,\lozenge}$, 
\setcounter{footnote}{0}Heng Hu$^{\P,\S,}$\footnote{Corresponding author. E-mail address: huheng@whu.edu.cn.}$^{}$
\end{center}
}

{\scriptsize{
\begin{center}
$^\P$School of Mathematics and Statistics, Ningxia University, 750021 Yinchuan, PR China\\
$^\S$School of Civil Engineering, Wuhan University, 8 South Road of East Lake, Wuchang, 430072 Wuhan, PR China\\
$^\ddagger$Institute of Physics, Chinese Academy of Sciences, Beijing 100180, China\\
$^\lozenge$Beijing Academy of Quantum Information Sciences, 100193 Beijing, China\\
$^\#$Universit\'{e} de Lorraine, CNRS, Arts et M\'{e}tiers ParisTech, LEM3, 57000, Metz, France
\end{center}
}}

\noindent\rule[-0.5ex]{16cm}{0.4pt}
\vspace{-1.1em}
\begin{flushleft}
\large\textbf{Abstract}
\end{flushleft}
\vspace{-1.1em}
\indent\indent Quantum computing offers a promising avenue for advancing computational methods in science and engineering. In this work, we introduce the quantum asymptotic numerical method (qANM), a framework for solving nonlinear path-following problems using quantum computing. Based on the principle of high-order perturbation techniques, the proposed method uses Taylor series expansions to transform complex nonlinear systems into sequences of linear equations, which are then solved using quantum linear solvers. The central objective of this study is to demonstrate nonlinear path-following with the required linear systems solved on real quantum hardware. To realize this objective, we develop the quantum-enhanced Jacobi method (q-Jacobi), an iterative quantum linear solver used in the hardware experiment. Numerical simulations on a quantum simulator validate the convergence of the method. A highlight of this work is a proof-of-principle experiment on a superconducting quantum processor. Despite the noise inherent in near-term quantum hardware, the experiment achieves 98\% accuracy in tracking the nonlinear solution path. We believe this work provides a useful reference for applying quantum computing to nonlinear computational mechanics.

\vspace{-1.1em}

\begin{flushleft}
\justifying\textbf{Keywords:} Asymptotic numerical method; Quantum computing; Perturbation method; Nonlinear problems.
\end{flushleft}
\vspace{-2.4em}
\noindent\rule[-0.5ex]{16cm}{0.4pt}

\section{Introduction}

Quantum computing represents a transformative paradigm in scientific computing, offering capabilities fundamentally different from those of classical computing \cite{song2019generation,arute2019quantum}. Unlike classical computers, which process information sequentially or in parallel across bits, quantum computers exploit principles such as superposition and entanglement to process vast amounts of information simultaneously \cite{grover2001schrodinger}. This unique potential has attracted increasing attention within the computational mechanics community, e.g., quantum computing has been applied to the finite element method \cite{ali2023performance,nguyen2024quantum,balducci2024solving,xu2025decomposition}, elastodynamics \cite{xu2026hamiltonian}, fluid dynamics \cite{meng2023quantum,schalkers2024efficient,meng2024simulating,lu2024quantum,Wang2026Simulating}, structural design \cite{wang2025quantum2,WULFF2024117380,wils2023symbolic,sukulthanasorn2025novel}, vibration \cite{wu2025quantum}, computational homogenization \cite{liu2024towards,xu2026towards}, and data-driven computing \cite{xu2024quantum}, among many others. Interested readers can refer to the recent review papers \cite{tosti2022review,moller2023quantum,meng2025challenges,zhaoyuan2025advances,yongchun2025quantum}. In this work, our focus is on investigating the potential usage of quantum computing in addressing nonlinear problems.

Nonlinear problems are widespread in science and engineering, with solid mechanics as a prime example, where materials exhibit complex behaviors such as nonlinear geometric deformations and stress-strain relationships \cite{trovalusci2003non,wriggers2008nonlinear,reddy2014introduction}. The Newton-Raphson (NR) method is perhaps the most widely used classical approach for solving such nonlinear problems \cite{de2012nonlinear}, but it faces two primary challenges. First, each NR iteration requires the solution of a linearized system, which can become computationally demanding for large-scale problems, especially in multiscale simulations involving vast numbers of finite elements \cite{han2020efficient}. Second, the NR method's reliance on local linear approximations limits its convergence region, often requiring multiple steps to track a solution path, or even failing to converge under strong nonlinearity conditions, as in shell buckling problems \cite{ferreira2000buckling,yang2023tutorial}. These challenges motivate nonlinear solution strategies that can reduce the number of continuation steps while maintaining stable path-following performance.

Quantum computing offers a promising direction to enhance the efficiency of solving linear equations, which arise from the linearization of the nonlinear problem. For instance, the Harrow-Hassidim-Lloyd (HHL) algorithm \cite{harrow2009quantum}, a quantum algorithm for linear systems, can generate a quantum solution state with an exponentially reduced computational cost $O(\log D)$. It has been applied to solve nonlinear differential equations \cite{leyton2008quantum}, dissipative nonlinear differential equations \cite{liu2021efficient}, and nonlinear algebraic equations \cite{xue2021quantum}. Additionally, the variational quantum linear solver (VQLS) \cite{bravo2023variational} is another promising method. By using a parameterized quantum circuit to approximate the solution to a linear equation, VQLS requires significantly fewer qubits and lower circuit depth than the HHL algorithm. It has been implemented for fluid dynamics simulations on a superconducting quantum processor \cite{chen2024enabling} and combined with the finite element method to solve the Poisson equation \cite{ali2023performance,trahan2023variational}. Nevertheless, developing a new quantum algorithm for solving linear equations remains an ongoing process \cite{pellow2023near,aaronson2015read,morales2024quantum}, and it may potentially handle the computational cost issue in nonlinear solvers.

Recent studies have begun to address nonlinear problems on quantum hardware. Nonlinear ground states have been investigated on IBM quantum processors using variational quantum algorithms \cite{lubasch2020variational,umer2025probing}. Selected expectation values required for a single time step of a nonlinear Black--Scholes equation have been evaluated on a trapped ion quantum processor \cite{sarma2024quantum}. Quantum finite difference schemes for time dependent nonlinear equations, including the Korteweg--De Vries equation, have also been verified on IBM quantum hardware \cite{chagneau2023quantum}. A different theoretical route converts a quadratic nonlinear system into an enlarged linear system before applying a quantum linear solver \cite{xue2022quantum}. These studies demonstrate the feasibility of several quantum approaches to nonlinear problems, while also showing present limitations associated with small problem sizes, encodings tailored to individual problems, hardware noise, and output measurement \cite{lubasch2020variational,umer2025probing,sarma2024quantum,chagneau2023quantum}. Their objectives are distinct from tracking a continuous equilibrium path in nonlinear mechanics, which is the focus of the present work.

Meanwhile, the asymptotic numerical method (ANM) \cite{cochelin2007method} offers the potential to expand the convergence region in solving nonlinear path-following problems. In many nonlinear mechanics problems, the goal is not only to compute a single equilibrium state, but also to trace the continuous evolution of the response under increasing loading or deformation. As a high-order perturbation technique, ANM describes this evolution by using Taylor series expansions, thereby transforming the original nonlinear equation into a sequence of linear equations. Compared with the local linear approximation used in the NR method, the Taylor series in ANM contains higher-order information about the nonlinear response. As a result, one ANM expansion can often cover a finite segment of the equilibrium path, reducing the number of nonlinear steps required in the computation. Numerous applications in solid and fluid mechanics have demonstrated ANM's efficiency and robustness \cite{potier2024asymptotic}. For instance, ANM has been used to predict the buckling of thin-walled structures \cite{huang2017fourier,zahrouni1999computing,kuang2021computational,hui2021hierarchical}, reduce computational costs in multiscale simulations \cite{nezamabadi2009multilevel,hui2019multiscale}, track instabilities in adhesive contact \cite{du2019asymptotic}, and analyze bifurcation in fluid dynamics \cite{guevel2018numerical}, among other applications.

In this study, we introduce the quantum asymptotic numerical method (qANM) for tracking nonlinear solution paths using quantum linear solvers. The central objective is to demonstrate that a nonlinear solution path can be tracked with the required linear systems solved on a real superconducting quantum processor. ANM constructs the continuous solution path and converts the nonlinear problem into a sequence of linear systems. The newly proposed quantum-enhanced Jacobi method (q-Jacobi) solves these systems without a variational optimization loop, and its iterative form allows convergence to be monitored directly. In addition, we provide a VQLS formulation for the numerical simulations to show that the qANM procedure is not restricted to q-Jacobi. We then present a proof-of-principle experiment on a superconducting quantum processor to demonstrate that the proposed qANM procedure can be implemented on current quantum hardware.

The remainder of this paper is organized as follows. \cref{Methodology} outlines the core qANM framework and the q-Jacobi method used in the numerical simulations presented in the main text and in the hardware experiment. \cref{sec: Validation} evaluates the quantum linear solvers and presents numerical validations using the spring-mass and Euler-Bernoulli beam buckling problems, including an investigation of solver inaccuracy and iterative refinement. \cref{sec: Experiment} details the proof-of-concept experiment on a superconducting quantum processor, followed by conclusions in \cref{sec: Conclusion}.

\section{Methodology}\label{Methodology}

In this section, we present the theoretical framework of the proposed qANM. ANM transforms the nonlinear problem into a sequence of linear systems, which are then solved using a quantum linear solver. \Cref{sec: ANM} presents the core ANM formulation, and \cref{sec: linear} presents the quantum-enhanced Jacobi method (q-Jacobi), which is used in the numerical investigations and in the experiment on a superconducting quantum processor.

\subsection{Linearization via the asymptotic numerical method}\label{sec: ANM}

We consider a nonlinear algebraic system dependent on a scalar parameter $\lambda$:
\begin{equation}\label{eq: Nonlinear}
\bm{R}(\bm{u}, \lambda) = 0,
\end{equation}
where $\bm{u} \in \mathbb{R}^D$ is the unknown vector of interest, and $\bm{R}: \mathbb{R}^D \times \mathbb{R} \rightarrow \mathbb{R}^D$ is a continuously differentiable function introducing nonlinearity into the system. 
The formulation in \cref{eq: Nonlinear} is widely common in scientific computing, appearing in contexts such as solid mechanics problems related to material and geometric nonlinearities~\cite{potier1997traitement, najah1998critical}, as well as the Navier-Stokes equations in fluid mechanics~\cite{guevel2018numerical, medale2009parallel}, among others.

The ANM is fundamentally a high-order perturbation technique. Historically, perturbation methods were developed to construct analytical approximate solutions for complex nonlinear differential equations by expanding them around a known state \cite{potier2024asymptotic,nayfeh2024perturbation,koiter1967stability}. The ANM adapts this classical mathematical concept into a robust numerical framework. Its primary objective is to compute a continuous solution path $a \mapsto (\bm{u}(a), \lambda(a))$ starting from a known solution point $(\bm{u}_0, \lambda_0)$. To achieve this, we introduce a perturbation parameter $a$ and express the unknown vector $\bm{u}$ and the parameter $\lambda$ as Taylor series expansions:
\begin{equation}\label{eq: taylor}
\begin{cases}
\bm{u}(a) = \bm{u}_0 + a \bm{u}_1 + a^2 \bm{u}_2 + \dots + a^N \bm{u}_N, \\
\lambda(a) = \lambda_0 + a \lambda_1 + a^2 \lambda_2 + \dots + a^N \lambda_N,
\end{cases}
\end{equation}
where $N \in \mathbb{Z}_{>0}$ denotes the order of the Taylor series, typically chosen as $N \leq 20$ for practical computations \cite{guillot2019generic}. To close the system of equations given by \cref{eq: Nonlinear,eq: taylor}, an additional constraint is required. The expansion parameter $a$ is defined through the arclength condition
\begin{equation}\label{eq: arclength_definition}
a=(\bm{u}(a)-\bm{u}_0)^\top\bm{u}_1+(\lambda(a)-\lambda_0)\lambda_1,
\end{equation}
which is commonly used in the Riks method in solid mechanics~\cite{potier2024asymptotic}.

Substituting the expansions from \cref{eq: taylor} into the nonlinear problem in \cref{eq: Nonlinear} gives a set of equations associated with different powers of the path parameter $a$. To see how these equations are obtained, we first measure the change of the unknowns from the current base point $(\bm{u}_0,\lambda_0)$. We define
\begin{equation}\label{eq: ANM_increment}
\Delta \bm{u}(a)=\bm{u}(a)-\bm{u}_0,
\qquad
\Delta \lambda(a)=\lambda(a)-\lambda_0,
\end{equation}
where $\Delta \bm{u}(a)$ and $\Delta \lambda(a)$ denote the displacement increment and the load-parameter increment along the local solution path. According to \cref{eq: taylor}, these increments are expressed as
\begin{equation}\label{eq: ANM_increment_series}
\Delta \bm{u}(a)=\sum_{p=1}^{N}a^p\bm{u}_p,
\qquad
\Delta \lambda(a)=\sum_{p=1}^{N}a^p\lambda_p .
\end{equation}
Here, $\bm{u}_p$ and $\lambda_p$ are the $p$-th order Taylor coefficients to be determined. They are not independent solutions at different load levels; rather, together they define a local polynomial representation of the equilibrium branch.

Since the expansion starts from a known equilibrium point, the residual satisfies
\begin{equation}\label{eq: base_equilibrium}
\bm{R}(\bm{u}_0,\lambda_0)=\bm{0}.
\end{equation}
Assuming that the residual is sufficiently smooth in a neighborhood of this point, we separate its first-order part from the remaining nonlinear terms:
\begin{equation}\label{eq: residual_expansion}
\bm{R}(\bm{u}_0+\Delta \bm{u},\lambda_0+\Delta \lambda)
=
\bm{K}\Delta \bm{u}-\bm{F}\Delta \lambda
+
\bm{R}_{\mathrm{nl}}(\Delta \bm{u},\Delta \lambda).
\end{equation}
In this expression,
\begin{equation}\label{eq: KF_definition}
\bm{K}=D_{\bm{u}}\bm{R}(\bm{u}_0,\lambda_0),
\qquad
\bm{F}=-D_{\lambda}\bm{R}(\bm{u}_0,\lambda_0).
\end{equation}
The matrix $\bm{K}\in\mathbb{R}^{D\times D}$ is the tangent matrix, or Jacobian matrix, obtained by differentiating the residual with respect to $\bm{u}$ at the current base point. The vector $\bm{F}\in\mathbb{R}^{D}$ is the load direction associated with the scalar parameter $\lambda$. The minus sign in the definition of $\bm{F}$ is introduced so that the first-order equilibrium equation takes the standard form $\bm{K}\bm{u}_1=\lambda_1\bm{F}$. The term $\bm{R}_{\mathrm{nl}}$ denotes the nonlinear remainder. It contains all contributions that are at least of second order with respect to the increments $\Delta\bm{u}$ and $\Delta\lambda$.

We now insert the series in \cref{eq: ANM_increment_series} into \cref{eq: residual_expansion}. The linear part gives
\begin{equation}\label{eq: linear_part_series}
\bm{K}\Delta\bm{u}(a)-\bm{F}\Delta\lambda(a)
=
\sum_{p=1}^{N}a^p
\left(
\bm{K}\bm{u}_p-\lambda_p\bm{F}
\right).
\end{equation}
The nonlinear remainder can also be written as a power series in $a$:
\begin{equation}\label{eq: nonlinear_part_series}
\bm{R}_{\mathrm{nl}}(\Delta\bm{u}(a),\Delta\lambda(a))
=
\sum_{p=2}^{N}a^p\bm{F}_{\mathrm{nl}}^{(p)}
+
O(a^{N+1}).
\end{equation}
The vector $\bm{F}_{\mathrm{nl}}^{(p)}\in\mathbb{R}^{D}$ is the coefficient of the nonlinear residual at order $p$. The summation starts from $p=2$ because $\bm{R}_{\mathrm{nl}}$ contains no constant term and no first-order term. The notation $O(a^{N+1})$ represents the terms beyond the truncation order $N$.

Combining \cref{eq: linear_part_series,eq: nonlinear_part_series}, the residual along the approximate path becomes
\begin{equation}\label{eq: residual_order_expansion}
\bm{R}(\bm{u}(a),\lambda(a))
=
\sum_{p=1}^{N}a^p
\left(
\bm{K}\bm{u}_p-\lambda_p\bm{F}
\right)
+
\sum_{p=2}^{N}a^p\bm{F}_{\mathrm{nl}}^{(p)}
+
O(a^{N+1}).
\end{equation}
For the polynomial path to satisfy the equilibrium equation up to order $N$, the coefficient of each power of $a$ must vanish. This gives the sequence of linear problems
\begin{equation}\label{eq: linear_sys}
\begin{cases}
\bm{K}\bm{u}_1=\lambda_1\bm{F},\\[2mm]
\bm{K}\bm{u}_p=\lambda_p\bm{F}-\bm{F}_{\mathrm{nl}}^{(p)},
\qquad 2\leq p\leq N.
\end{cases}
\end{equation}
This equation is the central algebraic structure of ANM. The same tangent matrix $\bm{K}$ appears at every order within the current continuation step, while the right-hand side changes according to the previously computed Taylor coefficients.

To keep the main derivation focused, the detailed construction of $\bm{F}_{\mathrm{nl}}^{(p)}$, the arclength conditions, and the recursive calculation of the Taylor coefficients are provided in \cref{sec: ANM_details}. The coefficient extraction is analytical: terms multiplying equal powers of $a$ are collected directly. When the problem at order $p$ is reached, the coefficients from orders $1$ to $p-1$ are already known. They determine $\bm{F}_{\mathrm{nl}}^{(p)}$, so the next linear system in \cref{eq: linear_sys} can be solved. The corresponding analytical form for the Euler--Bernoulli beam discretization is derived in \cref{sec: EB_ANM_coefficients}.

To determine the admissible range of $a$, ANM estimates the validity of the truncated series by comparing the last retained term with the first-order contribution. Requiring
\begin{equation}\label{eq: validity_condition}
\frac{\|a^N\bm{u}_N\|}{\|a\bm{u}_1\|}
\leq
\epsilon_d
\end{equation}
leads to the standard estimate of the validity radius
\begin{equation}\label{eq: amax}
a_{\mathrm{max}}
=
\left(
\epsilon_d
\frac{\|\bm{u}_1\|}{\|\bm{u}_N\|}
\right)^{\frac{1}{N-1}},
\end{equation}
where $\epsilon_d$ is a user-defined accuracy parameter and $\|\cdot\|$ denotes the Euclidean norm. The next point of the continuation procedure is then obtained by evaluating the polynomial solution at $a=a_{\mathrm{max}}$:
\begin{equation}\label{eq: continuation_update}
\bm{u}_{\mathrm{new}}
=
\bm{u}_0+\sum_{p=1}^{N}a_{\mathrm{max}}^p\bm{u}_p,
\qquad
\lambda_{\mathrm{new}}
=
\lambda_0+\sum_{p=1}^{N}a_{\mathrm{max}}^p\lambda_p.
\end{equation}
The pair $(\bm{u}_{\mathrm{new}},\lambda_{\mathrm{new}})$ becomes the base point of the next continuation step. Meanwhile, any intermediate value $a\in[0,a_{\mathrm{max}}]$ can be inserted into the Taylor representation in \cref{eq: taylor} to generate additional points on the same local path without resolving the nonlinear system. Thus, one ANM expansion describes a continuous path segment rather than only one equilibrium point.

The above construction gives one local ANM continuation step. Starting from a known equilibrium point $(\bm{u}_0,\lambda_0)$, the Taylor coefficients are obtained recursively by solving the linear systems in \cref{eq: linear_sys}, and the admissible interval of the local path parameter is determined by \cref{eq: amax}. The endpoint defined in \cref{eq: continuation_update} is then used as the base point for the next continuation step. Repeating this procedure allows the method to trace the solution branch until a prescribed stopping condition is reached, for example, when the load parameter $\lambda$ attains a target value.

The central difference from the Newton--Raphson method is that one ANM expansion represents a continuous segment of the equilibrium path rather than a correction at one prescribed load value. Within each continuation step, all Taylor orders use the same tangent matrix $\bm{K}$, and the validity radius $a_{\mathrm{max}}$ automatically limits the length of that segment. These features provide the path-following basis of qANM. A fuller explanation of the ANM construction and its interpretation is given in \cref{sec: ANM_details}, while the Newton--Raphson formulation is summarized in \cref{sec: NR}.

In summary, ANM provides the mathematical foundation of the proposed framework by converting the nonlinear system into a sequence of linear equations and representing a continuous segment of the solution path. These are the core ideas needed to understand qANM. Additional details of the coefficient construction are collected in \cref{sec: ANM_details}. The following subsection presents q-Jacobi, which is used in the numerical simulations presented in the main text and in the hardware experiment; the VQLS formulation used only in numerical simulations is given in \cref{sec: VQLS}.

\subsection{Quantum-enhanced Jacobi method for solving linear equations}\label{sec: linear}\label{sec: Jacobi}

In this subsection, we present the q-Jacobi method used to solve the linear systems obtained from the ANM expansion in \cref{eq: linear_sys}. The role of ANM is to provide the nonlinear-to-linear transformation, while each resulting linear system can, in principle, be solved by different linear solvers. The main text therefore focuses on q-Jacobi because it is the method implemented on the superconducting quantum processor. As a supplementary quantum linear solver, VQLS is included only in the numerical simulations and is described in \cref{sec: VQLS}. For clarity, we write a representative linear system in the form
\begin{equation}\label{eq: KuF}
\bm{K}\bm{u}=\bm{F},
\end{equation}
where $\bm{K}\in\mathbb{R}^{D\times D}$ is the system matrix, $\bm{u}\in\mathbb{R}^{D}$ is the unknown vector, and $\bm{F}\in\mathbb{R}^{D}$ is the right-hand side vector. In the ANM context, $\bm{K}$ corresponds to the tangent matrix at the current continuation point, while $\bm{F}$ represents a generic right-hand side, including the load direction or the higher-order nonlinear terms generated by the Taylor expansion.

The q-Jacobi method developed in this work is based on the classical Jacobi iteration~\cite{saad2003iterative}, but uses a quantum subroutine to estimate the inner products required in the matrix-vector multiplication. It does not require a variational optimization loop, and its iterative form allows convergence to be monitored directly through the residual. These features make q-Jacobi the main quantum algorithmic development used in the hardware experiment.

The classical Jacobi method solves the linear system in \cref{eq: KuF} iteratively by decomposing the matrix $\bm{K}$ into its diagonal component $\bm{A}$ and the remainder $\bm{T}$:
\begin{equation}\label{eq: jacobi_decompose}
\bm{K} = \bm{A} + \bm{T},
\end{equation}
where $\bm{A} \in \mathbb{R}^{D \times D}$ is a diagonal matrix containing the diagonal elements of $\bm{K}$, and $\bm{T} = \bm{K} - \bm{A}$ contains the off-diagonal elements. The iterative formula of the Jacobi method is then given by
\begin{equation}\label{eq: jacobi_iteration}
\bm{u}^{(k+1)} = \bm{A}^{-1} (\bm{F} - \bm{T} \bm{u}^{(k)}) = -\bm{A}^{-1} \bm{T} \bm{u}^{(k)} + \bm{A}^{-1} \bm{F}.
\end{equation}
To simplify the notation, we define
\begin{equation}\label{eq: jacobi_Gc}
\bm{M} = -\bm{A}^{-1} \bm{T}, \quad \bm{c} = \bm{A}^{-1} \bm{F}.
\end{equation}
Then the iteration becomes
\begin{equation}\label{eq: jacobi_iteration_simplified}
\bm{u}^{(k+1)} = \bm{M} \bm{u}^{(k)} + \bm{c}.
\end{equation}
The iteration process in \cref{eq: jacobi_iteration_simplified} takes the majority of the total cost of the Jacobi method. More specifically, the computational complexity for matrix-vector multiplication, i.e., $\bm{M} \bm{u}^{(k)}$, is generally $O(D^2)$ on a classical computer. After obtaining $\bm{M} \bm{u}^{(k)}$, the rest of the computation of adding $\bm{c}$ is only $O(D)$. Therefore, the complexity of one iteration is $O(D^2)$ on a classical computer \cite{wendland2017numerical}.

\begin{figure}[!htbp]
\centering
\includegraphics[width=0.5\textwidth]{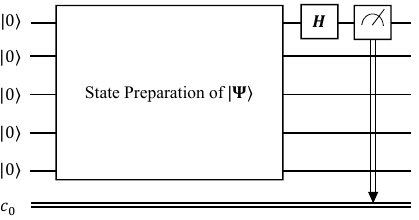}
\caption{Quantum circuit for computing the inner product between a row of $\bm{M}$ and the vector $\bm{u}^{(k)}$ using the H-based algorithm. The circuit involves state preparation of $\ket{\bm{\Psi}}$, application of the Hadamard gate $\bm{H}$, and measurement of the ancilla qubit.}
\label{fig: qJacobi_Hadamard}
\end{figure}

To reduce the complexity of the matrix-vector product $\bm{M} \bm{u}^{(k)}$, the q-Jacobi method uses an H-based quantum algorithm to perform inner product calculations \cite{kuang2025quantum,zhao2021compiling,moradi2022clinical}, as shown in  \cref{fig: qJacobi_Hadamard}.
To compute each component of $\bm{M} \bm{u}^{(k)}$, we proceed as follows. Let $\bm{m}_i$ denote the $i$-th row of the matrix $\bm{M}$, we normalize $\bm{m}_i$ and the vector $\bm{u}^{(k)}$:
\begin{equation}\label{eq: normalization}
\tilde{\bm{m}}_i = \frac{\bm{m}_i}{\| \bm{m}_i \|}, \quad \tilde{\bm{u}}^{(k)} = \frac{\bm{u}^{(k)}}{\| \bm{u}^{(k)} \|},
\end{equation}
where $\| \cdot \|$ denotes the Euclidean norm. We then prepare the quantum state
\begin{equation}\label{eq: Phi_state}
|\bm{\Psi}\rangle = \frac{1}{\sqrt{2}} \left( |0\rangle |\tilde{\bm{m}}_i\rangle + |1\rangle |\tilde{\bm{u}}^{(k)}\rangle \right),
\end{equation}
where $|\tilde{\bm{m}}_i\rangle$ and $|\tilde{\bm{u}}^{(k)}\rangle$ are quantum states encoding the normalized vectors $\tilde{\bm{m}}_i$ and $\tilde{\bm{u}}^{(k)}$, respectively. The representation of the quantum state $|\bm{\Psi}\rangle$ requires $\log_2 2D = n_q +1$ qubits, matching the qubit count of the VQLS formulation in \cref{sec: VQLS}. Similar to the assumption used in VQLS \cite{bravo2023variational} and HHL \cite{harrow2009quantum}, we assume the state $|\bm{\Psi}\rangle$ can be efficiently prepared with a complexity of $O(\mathrm{polylog} (D))$, like using the efficient state preparation algorithms \cite{nakaji2022approximate,marin2023quantum,zylberman2024efficient} or quantum device \cite{giovannetti2008quantum}.

Applying the Hadamard gate $\bm{H}$ to the ancilla qubit of the state $|\bm{\Psi}\rangle$, we obtain
\begin{align}\label{eq: after_Hadamard}
\bm{H} |\bm{\Psi}\rangle &= \frac{1}{\sqrt{2}} \left( \bm{H} |0\rangle |\tilde{\bm{m}}_i\rangle + \bm{H} |1\rangle |\tilde{\bm{u}}^{(k)}\rangle \right) \notag \\
&= \frac{1}{\sqrt{2}} \left( \frac{|0\rangle + |1\rangle}{\sqrt{2}} |\tilde{\bm{m}}_i\rangle + \frac{|0\rangle - |1\rangle}{\sqrt{2}} |\tilde{\bm{u}}^{(k)}\rangle \right) \notag \\
&= \frac{1}{2} \left( |0\rangle (|\tilde{\bm{m}}_i\rangle + |\tilde{\bm{u}}^{(k)}\rangle) + |1\rangle (|\tilde{\bm{m}}_i\rangle - |\tilde{\bm{u}}^{(k)}\rangle) \right).
\end{align}
The probability $P_0$ of measuring the ancilla qubit in the state $|0\rangle$ is given by the squared norm of the component of $\bm{H} |\bm{\Psi}\rangle$ corresponding to $|0\rangle$:
\begin{equation}\label{eq: P0_derivation}
P_0 = \left\| \frac{1}{2} \left( |\tilde{\bm{m}}_i\rangle + |\tilde{\bm{u}}^{(k)}\rangle \right) \right\|^2 = \frac{1}{2} + \frac{1}{2} \langle \tilde{\bm{m}}_i | \tilde{\bm{u}}^{(k)} \rangle.
\end{equation}
Therefore, we can estimate the inner product as
\begin{equation}\label{eq: inner_product_estimate}
\langle \tilde{\bm{m}}_i | \tilde{\bm{u}}^{(k)} \rangle = 2 P_0 - 1.
\end{equation}
Each component of the matrix-vector product $\bm{M} \bm{u}^{(k)}$ is then recovered using the relation
\vspace{-5mm}
\begin{equation}\label{eq: component_recovery}
(\bm{M} \bm{u}^{(k)})_i = \| \bm{m}_i \| \| \bm{u}^{(k)} \| \langle \tilde{\bm{m}}_i | \tilde{\bm{u}}^{(k)} \rangle.
\end{equation}
This computation repeats for each row of $\bm{M}$, i.e., $\bm{m}_i$ $(\text{for } i=1,2,..., D)$, then one can finish the matrix-vector multiplication $\bm{M} \bm{u}^{(k)}$ on a quantum computer, thereby performing the Jacobi iteration in \cref{eq: jacobi_iteration_simplified} until
\begin{equation}\label{eq: Tol}
Tol = \| \bm{u}^{(k)} - \bm{u}^{(k-1)} \| / \| \bm{u}^{(k-1)} \| < \epsilon_J,
\end{equation}
where $\epsilon_J$ is a user-defined tolerance as a terminal condition. In this way, the quantum solution of \cref{eq: KuF} is obtained as $\bm{u}=\bm{u}^{(k)}$. The complete q-Jacobi method is outlined in \cref{alg: qJacobi}. 

\begin{algorithm}[t]
\small
\caption{Quantum-enhanced Jacobi method (q-Jacobi)}
\begin{algorithmic}[1]
\Require System matrix $\bm{K} \in \mathbb{R}^{D \times D}$, right-hand side vector $\bm{F} \in \mathbb{R}^D$, initial guess $\bm{u}^{(0)}$, tolerance $\epsilon_J$.
\State \textbf{Preprocessing:} Prepare $\bm{M}$ and $\bm{c}$ in \cref{eq: jacobi_Gc}, normalize each row $\bm{m}_i$ of $\bm{M}$, store $\tilde{\bm{m}}_i$ and $\| \bm{m}_i \|$, set $k = 0$.
\Repeat
    \State Normalize $\bm{u}^{(k)}$ to obtain $\tilde{\bm{u}}^{(k)}$ and store $\| \bm{u}^{(k)} \|$.
    \For{$i = 1$ to $D$}
        \State Prepare the state $|\bm{\Psi}\rangle$ as in equation~\cref{eq: Phi_state}.
        \State Apply the Hadamard gate to the ancilla qubit.
        \State Measure the ancilla qubit to estimate $P_0$.
        \State Compute $\langle \tilde{\bm{m}}_i | \tilde{\bm{u}}^{(k)} \rangle = 2 P_0 - 1$.
        \State Compute $(\bm{M} \bm{u}^{(k)})_i$ using equation~\cref{eq: component_recovery}.
    \EndFor
    \State Update $\bm{u}^{(k+1)} = \bm{M} \bm{u}^{(k)} + \bm{c}$.
    \State $k \leftarrow k + 1$.
\Until{$Tol=\| \bm{u}^{(k)} - \bm{u}^{(k-1)} \| / \| \bm{u}^{(k-1)} \| < \epsilon_J$}
\State \Return Solution vector $\bm{u}^{(k)}$.
\end{algorithmic}
\label{alg: qJacobi}
\end{algorithm}

\begin{table}[t]
  \centering
  \footnotesize
  \caption{Computational complexity of the q-Jacobi method per iteration. Note that the complexity $O(\mathrm{polylog} (D))$ for preparing $\ket{\bm{\Psi}}$ is assumed here using the efficient state preparation methods in \cite{nakaji2022approximate,marin2023quantum,zylberman2024efficient,zhang2022quantum}.}
    \begin{tabular}{cccc}
    \toprule
    Procedures & Complexity & Repetitions & Total complexity \\
    \midrule
    Normalizing $\bm{u}^{(k)}$ & $O(D)$ & 1 & \multirow{4}{*}{$O(D\mathrm{polylog} (D))$} \\
    \cmidrule{1-3}
    State preparation of $\ket{\bm{\Psi}}$ & $O(\mathrm{polylog} (D))$ & $D$ & \\
    \cmidrule{1-3}
    Gate and measurement & $O(1)$ & $D$ & \\
    \cmidrule{1-3}
    Adding $\bm{c}$ & $O(D)$ & 1 & \\
    \bottomrule
    \end{tabular}\label{table: qjacobi_complexity}\end{table}

The computational complexity of the proposed q-Jacobi method can be divided into two parts: preprocessing and iterative updates. In the preprocessing stage, the iteration matrix $\bm{M}=-\bm{A}^{-1}\bm{T}$ is constructed from $\bm{K}$, and each row $\bm{m}_i$ is normalized and stored together with its norm $\|\bm{m}_i\|$. Both the assembly of $\bm{K}$ and the construction of $\bm{M}$ are performed on a classical computer. For a fixed $\bm{K}$, this preprocessing has a one-time cost of $O(D^2)$ for a dense matrix representation. During each Jacobi iteration, the vector $\bm{u}^{(k)}$ is first normalized with a classical cost of $O(D)$. Then, for each row $\bm{m}_i$ of $\bm{M}$, the quantum circuit in \cref{fig: qJacobi_Hadamard} is used to estimate the inner product $\langle \tilde{\bm{m}}_i|\tilde{\bm{u}}^{(k)}\rangle$. Under the assumption that the state $\ket{\bm{\Psi}}$ in \cref{eq: Phi_state} can be prepared with complexity $O(\mathrm{polylog}(D))$ using efficient state-preparation methods~\cite{nakaji2022approximate,marin2023quantum,zylberman2024efficient,zhang2022quantum}, the $D$ inner-product evaluations lead to a per-iteration cost of $O(D\mathrm{polylog}(D))$, as summarized in \cref{table: qjacobi_complexity}. The remaining vector update, including the addition of $\bm{c}$, has a cost of $O(D)$ and does not change the leading scaling. Therefore, under the above state-preparation assumption and before accounting for the number of measurement shots, the per-iteration complexity of q-Jacobi is $O(D\mathrm{polylog}(D))$. If a conventional state-preparation procedure is used, preparing a state with $D$ amplitudes may require $O(D)$ operations in the worst case~\cite{mottonen2004transformation,iten2016quantum}, in which case this advantage over the classical Jacobi matrix-vector multiplication with cost $O(D^2)$ becomes much weaker. In addition, estimating each probability requires repeated circuit executions, so the shot cost must be included when assessing the end-to-end computational cost on hardware.

When q-Jacobi is incorporated into qANM, the same tangent matrix $\bm{K}$ is reused for all Taylor orders within one continuation step. Thus, although $N$ linear systems with different right-hand sides are solved in that step, $\bm{M}$ is constructed only once and reused for all Taylor orders. The preprocessing is repeated once for each new continuation step. If $N$ linear systems are solved in one qANM step and $k$ denotes a representative number of q-Jacobi iterations required for each linear solve, the leading cost associated with q-Jacobi can be written as
\begin{equation}
O\!\left(D^2 + kND\mathrm{polylog}(D)\right),
\end{equation}
under the same state-preparation assumptions. Since the Taylor order $N$ is typically moderate in practical ANM computations~\cite{guillot2019generic}, this structure is attractive from the viewpoint of reusing a fixed operator within each continuation step. However, this estimate should be interpreted with care. It does not include the cost of assembling the nonlinear right-hand sides $\bm{F}_{\mathrm{nl}}^{(p)}$, preparing the corresponding quantum states, performing repeated measurements, or reading out the solution vectors required by the ANM recursion. Moreover, the comparison here is only with the classical Jacobi matrix-vector multiplication, not with optimized sparse direct solvers, Krylov methods, or multigrid methods that are widely used for structured systems in computational mechanics. Therefore, the complexity estimate above applies only to the q-Jacobi quantum subroutine and does not establish an advantage over classical solvers. The hardware experiment in \cref{sec: Experiment} examines its behavior on a current quantum device.

Quantum state preparation is a widely recognized and unresolved challenge in quantum computing, rather than a difficulty specific to q-Jacobi. Quantum algorithms for linear algebra and differential equations, including HHL~\cite{harrow2009quantum,aaronson2015read}, QSVT \cite{gilyen2019quantum}, and many quantum machine learning algorithms~\cite{biamonte2017quantum} all face this input bottleneck; state preparation is therefore a central part of their end-to-end complexity~\cite{lin2024quantum}. The fact that this bottleneck is widespread does not lessen its importance. On the contrary, it makes efficient state preparation a fundamental problem for practical quantum computing. It also remains an active research direction, with continuing efforts on approximate and structured state preparation algorithms~\cite{nakaji2022approximate,marin2023quantum,zylberman2024efficient} and hardware approaches to coherent memory access. For example, Wang et al.~\cite{wang2025quantum} recently proposed a QRAM architecture based on transmon-controlled phonon routing. These developments provide possible directions for reducing the state preparation bottleneck and are relevant to the future practical implementation of the proposed method.

Finally, we clarify the role of the number of shots, denoted by $n_s$, in the quantum computations. In a quantum circuit, each measurement produces a classical outcome sampled from the probability distribution of the final quantum state. Therefore, a single circuit execution does not directly provide the probability itself. To estimate a quantity such as the probability $P_0$ in \cref{eq: P0_derivation}, the circuit must be executed repeatedly. If the circuit is run for $n_s$ shots and $n_0$ measurements return the outcome $0$, then $P_0$ is estimated as $P_0=n_0/n_s$. The statistical error of this estimate decreases as the number of shots increases, but the total number of circuit executions also increases accordingly. Thus, $n_s$ is an important part of the practical computational cost of qANM on quantum hardware. Its influence on the accuracy and convergence of the q-Jacobi solver will be examined in \cref{sec: Validation}.

In summary, qANM tracks nonlinear solution paths through the high-order ANM formulation, with the resulting linear systems solved by quantum linear solvers. In this framework, ANM converts the nonlinear problem into a sequence of linear systems, and the quantum component is used only for solving these systems within the continuation procedure. The robustness in tracking nonlinear solution paths mainly comes from the ANM formulation rather than from the choice of a particular quantum linear solver. The VQLS formulation in \cref{sec: VQLS} is used only in numerical simulations to illustrate the solver-independent nature of qANM. The main text focuses on q-Jacobi because its training-free and simpler circuit structure enables the small-scale implementation on real quantum hardware. The practical efficiency of the quantum part remains conditional on state preparation, matrix encoding, measurement cost, and output extraction. The following sections validate this framework through numerical simulations and a proof-of-principle experiment on superconducting quantum hardware.

\section{Validation}\label{sec: Validation}

In this section, we perform numerical tests on the quantum simulator Qiskit to evaluate q-Jacobi and the proposed qANM for nonlinear problems. The supplementary VQLS implementation used in the numerical simulations is described in \cref{sec: VQLS}. The validation in the main text focuses on the spring-mass problem used in the hardware experiment and on the influence of solver inaccuracy and iterative refinement in Euler-Bernoulli beam buckling. The analytical ANM coefficient construction for the beam buckling problem is provided in \cref{sec: EB_ANM_coefficients}. An additional beam deflection example is provided in \cref{sec: Numerical}.

\subsection{Evaluation of the quantum linear solvers}\label{sec: Evaluation}

To assess the effectiveness of VQLS and q-Jacobi in solving linear equations, we consider a set of linear systems $\bm{K} \bm{u} = \bm{F}_j$, where
\begin{equation}\label{eq: K2by2}
\bm{K} = \begin{bmatrix}
2 & -1 \\
-1 & 2 
\end{bmatrix}, \quad \bm{F}_j = \begin{bmatrix}
\cos\left(\dfrac{\pi}{4} j\right) \\
\sin\left(\dfrac{\pi}{4} j\right)
\end{bmatrix}, \quad \text{with } j = 0,1,\dots,7.
\end{equation}
The matrix $\bm{K}$ is chosen as a representative example in engineering numerical analysis~\cite{strang2007computational}, and the vectors $\bm{F}_j$ correspond to eight points sampled uniformly around the unit circle, providing multiple test cases. To evaluate the accuracy of the solutions $\bm{u}$ obtained from the quantum linear solvers, we use the following metric:
\begin{equation}\label{Acc}
\text{Accuracy} = \left(1 - \dfrac{\| \bm{u} - \bm{u}_{\text{ref}} \|}{\| \bm{u}_{\text{ref}} \|} \right) \times 100\%,
\end{equation}
where $\bm{u}_{\text{ref}}$ is the reference solution computed classically. It is important to note that for the q-Jacobi method, the solution vector is updated classically, avoiding global phase ambiguity. For VQLS, the global phase reduces to a sign ambiguity ($\pm 1$) in the real domain, which is automatically corrected by the scale factor $s$ during the reconstruction of $\bm{u}$, as detailed in \cref{sec: VQLS}. Thus, the metric in \cref{Acc} is compatible with both methods.

We detail the numerical setup as follows. For VQLS, we use 2 qubits and use the COBYLA optimizer, a gradient-free method suitable for variational quantum computing \cite{powell1998direct,bonet2023performance}. The ansatz consists of one layer of the hardware-efficient circuit shown in \cref{fig: VQLS}(a). The matrix $\bm{K}$ is decomposed into a linear combination of unitary matrices using the method described in~\cite{pesce2021h2zixy}. The unitary operator $\bm{U}$ for encoding $\bm{F}_j$ is obtained from a state preparation circuit generated by the \texttt{prepare\_state} method of \texttt{QuantumCircuit} in Qiskit~\cite{shende2004synthesis}.
For q-Jacobi, we also use 2 qubits and adopt the weighted Jacobi iteration to improve convergence, introducing a relaxation factor $\omega = 2/3$ in the iteration formula~\cite{saad2003iterative}. The state $\ket{\bm{\Psi}}$ in \cref{eq: Phi_state} is prepared using the \texttt{prepare\_state} method of \texttt{QuantumCircuit} in Qiskit. Following Shende et al.~\cite{shende2004synthesis}, the method recursively constructs the preparation circuit using multiplexed $\bm{R}_z$ and $\bm{R}_y$ rotations, which are decomposed into single-qubit rotations and CNOT gates.
Simulations are performed using the \textit{statevector} backend in Qiskit. To efficiently simulate the statistical estimation of probabilities required in VQLS and q-Jacobi, we use a normal approximation to accelerate the simulation of measurement statistics under a large number of shots, while retaining the effect of shot noise~\cite{kuang2025quantum,li2017efficient}.

\begin{figure}[!tb]
\centering
\includegraphics[width=\textwidth]{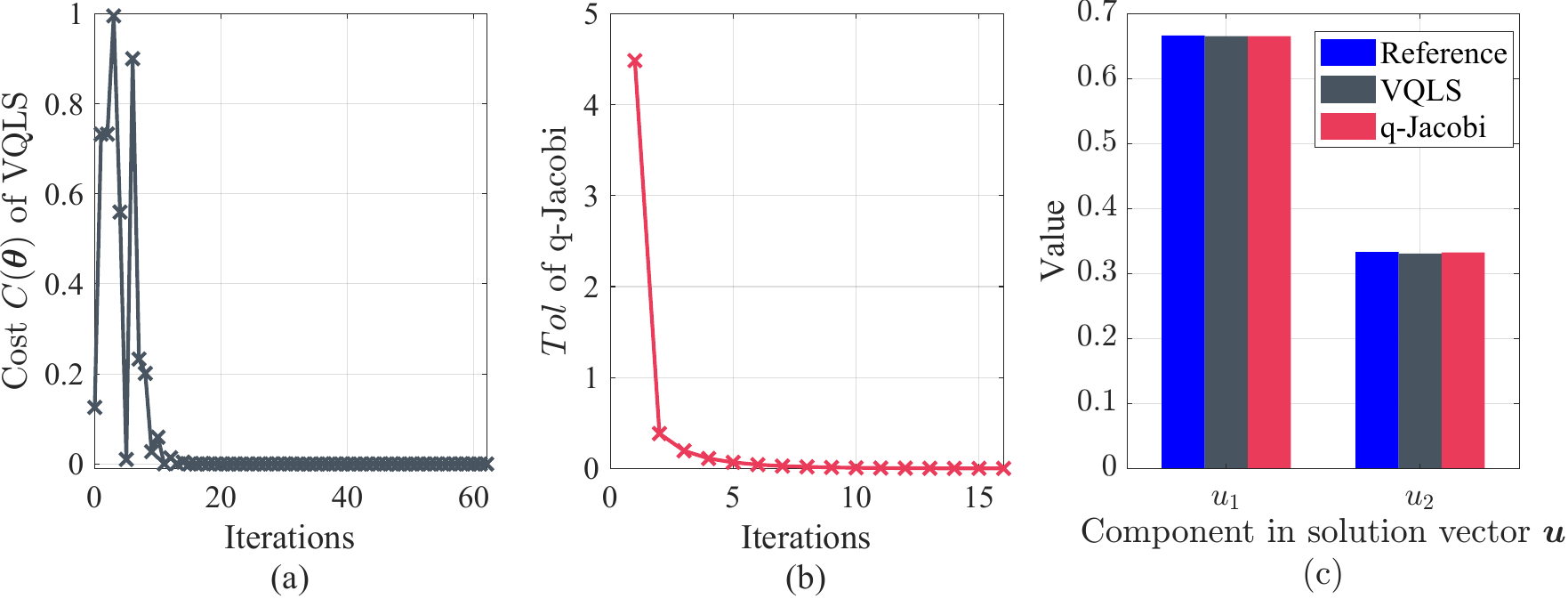}
\caption{Performance comparison of VQLS and q-Jacobi in solving $\bm{K} \bm{u} = \bm{F}_0$. (a) Cost function $C(\bm{\theta})$ versus iterations in VQLS. (b) Tolerance $Tol$ versus iterations in q-Jacobi. (c) Solutions $\bm{u}$ obtained by VQLS and q-Jacobi compared to the reference solution.}
\label{fig: iter_and_u}
\end{figure}

\begin{figure}[!tb]
\centering
\includegraphics[width=0.5\textwidth]{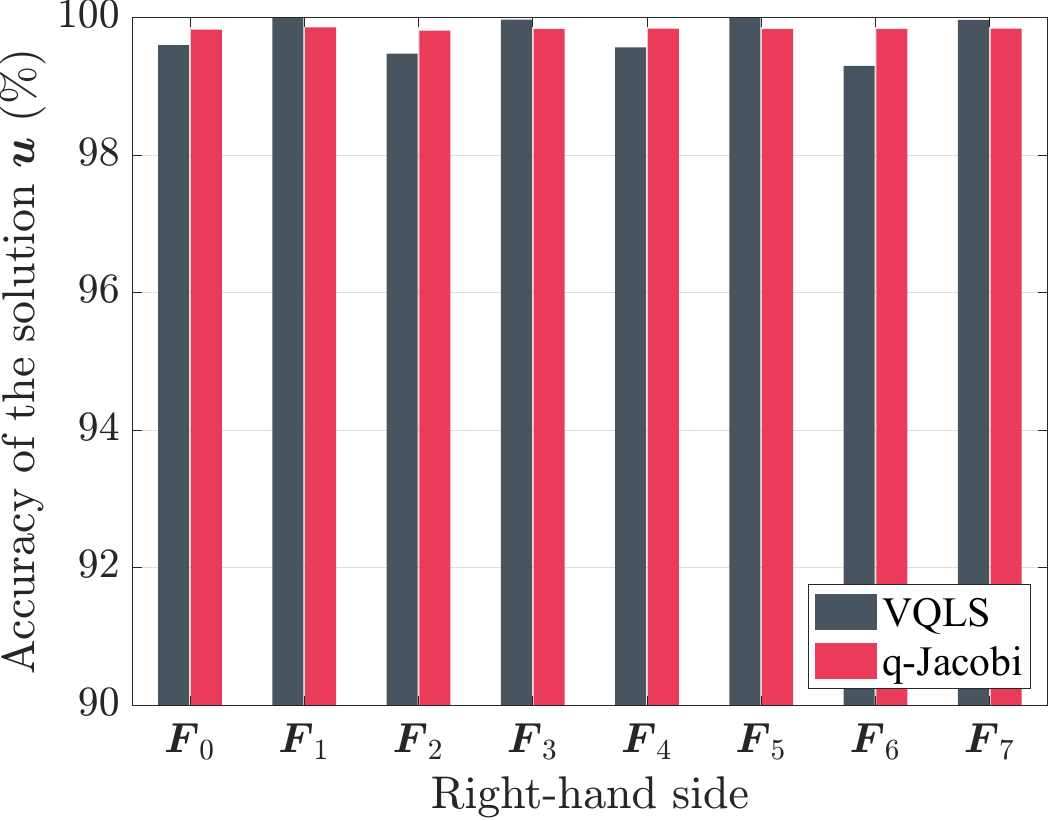}
\caption{Accuracy of solutions obtained by VQLS and q-Jacobi for different right-hand side vectors $\bm{F}_j$.}
\label{fig: F_j_and_u_results}
\end{figure}

\Cref{fig: iter_and_u} presents the results for solving $\bm{K} \bm{u} = \bm{F}_0$ using both VQLS and q-Jacobi, with the number of shots $n_s$ set to $10^8$. \Cref{fig: iter_and_u}(a) shows the convergence of the cost function $C(\bm{\theta})$ in VQLS over optimization iterations, while \Cref{fig: iter_and_u}(b) depicts the decrease of the tolerance $Tol$ in q-Jacobi iterations. Both algorithms demonstrate convergence after a certain number of iterations. The solutions $\bm{u}$ obtained from both algorithms are shown in \Cref{fig: iter_and_u}(c), along with the reference solution for comparison. The accuracies achieved are $99.63\%$ for VQLS and $99.88\%$ for q-Jacobi, indicating that both quantum algorithms can accurately solve the linear system. To further assess the performance of the algorithms, \Cref{fig: F_j_and_u_results} displays the accuracies of the solutions obtained by VQLS and q-Jacobi for each $\bm{F}_j$. Each bar represents the average accuracy over 10 runs to ensure statistical significance. Both algorithms consistently achieve accuracies greater than $99\%$ across all test cases, confirming their effectiveness in solving linear systems.

\begin{figure}[!t]
\centering
\includegraphics[width=0.8\textwidth]{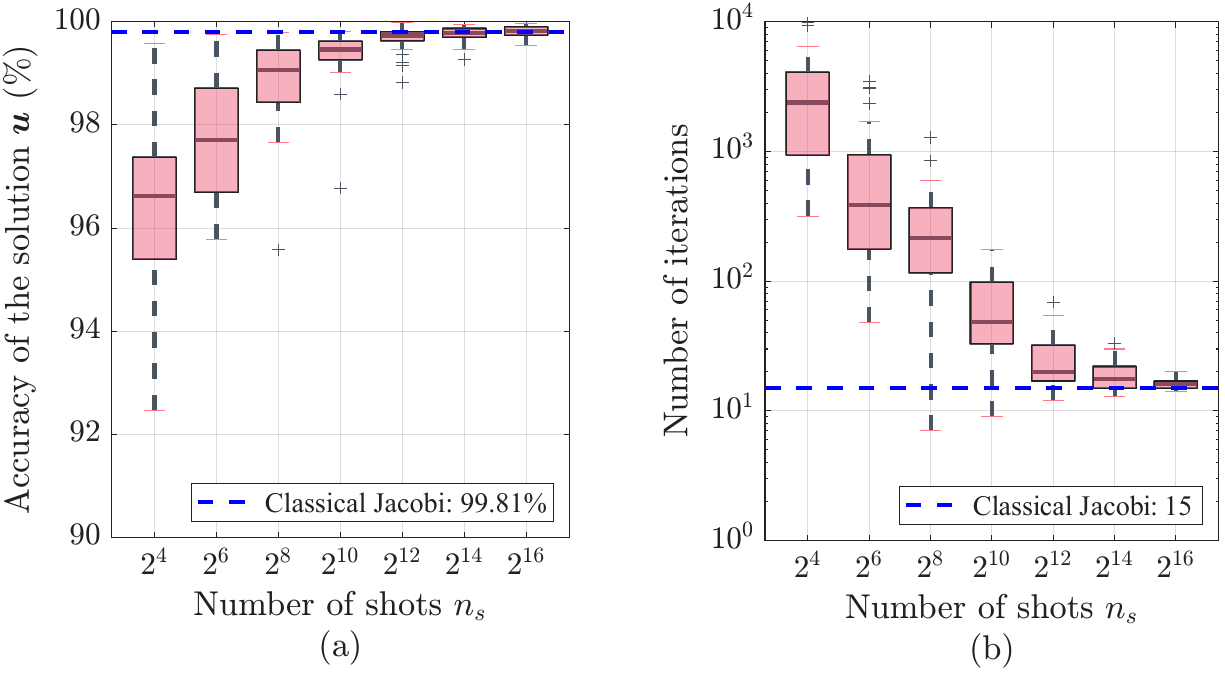}
\caption{Impact of the number of shots $n_s$ on the performance of q-Jacobi for solving $\bm{K} \bm{u} = \bm{F}_0$. (a) Accuracy of the solution versus $n_s$, with the classical Jacobi method accuracy as a reference (blue dashed line). (b) Number of iterations required for convergence versus $n_s$, compared to the classical Jacobi method (blue dashed line).}
\label{fig: shots_q_Jacobi}
\end{figure}

Moreover, we investigate the influence of the number of shots $n_s$ on the performance of the two quantum linear solvers. For VQLS, existing studies have already examined this influence, and interested readers can refer to \cite{bravo2023variational,trahan2023variational} for details. Here, we focus on the proposed q-Jacobi. \Cref{fig: shots_q_Jacobi}(a) shows the accuracy of the solution $\bm{u}$ versus $n_s$ when solving $\bm{K} \bm{u} = \bm{F}_0$, with the accuracy of the classical Jacobi method indicated by a blue dashed line for reference. Each box represents the data from 30 runs. As $n_s$ increases, the accuracy of q-Jacobi approaches that of the classical Jacobi method. Similarly, \Cref{fig: shots_q_Jacobi}(b) shows that the number of iterations required for convergence in q-Jacobi decreases with increasing $n_s$, converging toward the classical Jacobi method's iteration count. 
These results stem from the statistical nature of quantum measurements, which produce probabilistic outcomes that require multiple shots, as mentioned in \cref{sec: Jacobi}. Specifically, each shot contributes to estimating the probability $P_0$ in \cref{eq: P0_derivation}, and the estimation error in $P_0$ scales as $O(1/\sqrt{n_s})$~\cite{uno2021modified,suzuki2020amplitude}. Consequently, as $n_s$ increases, the accuracy of $P_0$ improves, leading to a more precise calculation of the matrix-vector product $\bm{M} \bm{u}^{(k)}$. This enhanced precision allows q-Jacobi to closely match the performance of the classical Jacobi method. 

\begin{figure}[t]
\centering
\includegraphics[width=0.5\textwidth]{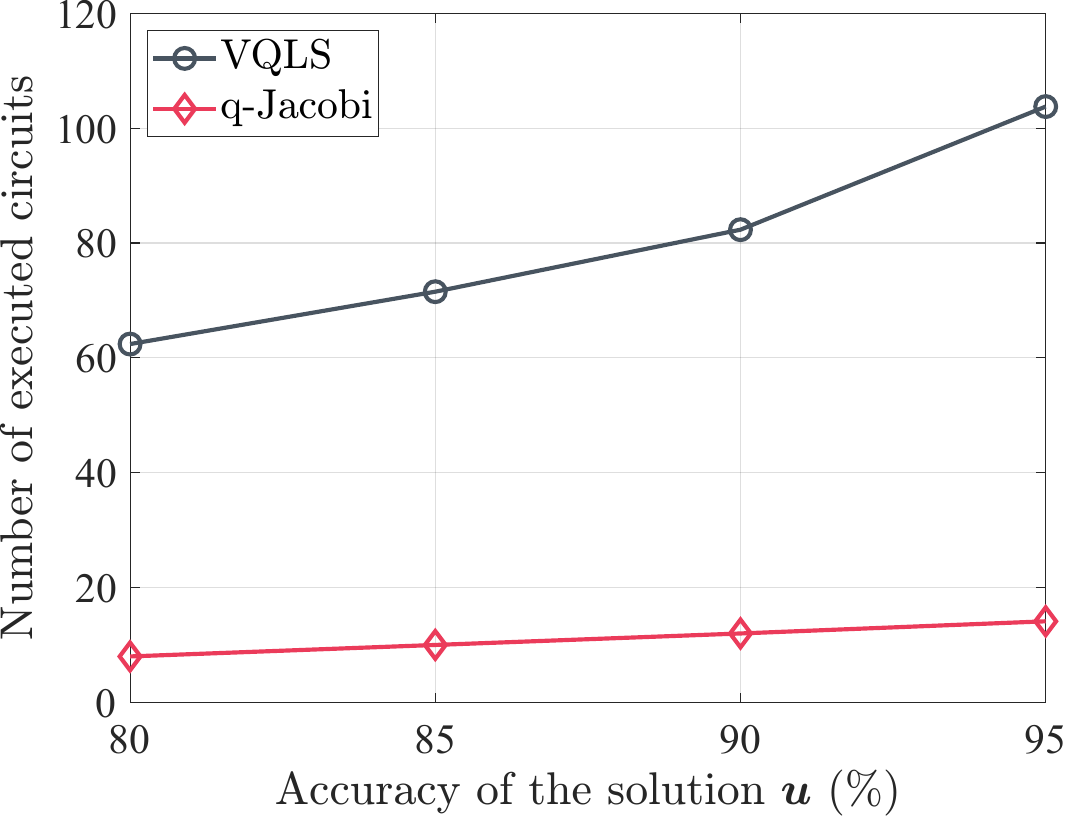}
\caption{Comparison of VQLS and q-Jacobi in terms of the number of executed quantum circuits required to achieve a given solution accuracy for $\bm{K} \bm{u} = \bm{F}_0$.}
\label{fig: vqls_Jacobi_comparsion}
\end{figure}

Finally, we compare VQLS and q-Jacobi from the perspective of hardware implementation. This comparison is not intended to establish a general superiority of one quantum linear solver over the other, but to motivate the choice of the solver used in the hardware experiment in \cref{sec: Experiment}. The linear equation $\bm{K}\bm{u}=\bm{F}_0$ is solved, and the number of shots is set to $n_s=10^8$ for both algorithms, meaning that each circuit execution consists of $n_s$ repetitions. \Cref{fig: vqls_Jacobi_comparsion} plots the solution accuracy versus the minimum number of executed circuits for both algorithms, with each point averaged over 100 samples.

The results show that, in this specific low-dimensional test, q-Jacobi reaches a given accuracy with fewer executed circuits than VQLS. For example, to achieve a solution accuracy of $95\%$, q-Jacobi requires approximately 14 circuit executions, while VQLS requires around 100 circuit executions. This behavior is mainly due to the training-free nature of q-Jacobi. Unlike VQLS, it does not require a variational optimization loop, whose convergence depends on the ansatz, the optimizer, the initial parameters, and the noise accumulated during repeated circuit evaluations~\cite{pellow2021comparison}. Therefore, although q-Jacobi should not be interpreted as outperforming mature classical sparse linear solvers, or as having a better asymptotic scaling than VQLS, it is more convenient for the small-scale real-device experiment considered in this work. For this reason, q-Jacobi is selected as the quantum linear-solver component in the superconducting quantum processor experiment presented in \cref{sec: Experiment}.

\subsection{The spring-mass problem}\label{sec: spring}

In this subsection, we validate the effectiveness of the proposed qANM by solving a spring-mass problem~\cite{cochelin2007method}.

\begin{figure}[!t]
\centering
\includegraphics[width=9cm]{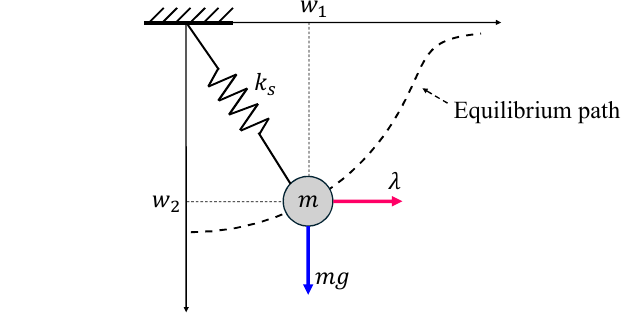}
\caption{Schematic of the spring-mass problem~\cite{cochelin2007method}.}
\label{fig: spring_ball}
\end{figure}

\Cref{fig: spring_ball} illustrates the spring-mass system, where a ball is subjected to three forces: gravity, the tension of the spring, and an applied force $\lambda$. The governing equations of this system are:
\begin{subequations}\label{eq: spring}
\begin{align}
& k_s (l - l_0) \frac{\partial l}{\partial w_1} - \lambda = 0,  \\
& k_s (l - l_0) \frac{\partial l}{\partial w_2} - mg = 0,
\end{align}
\end{subequations}
where $w_1$ and $w_2$ are the positions of the ball, $l = \sqrt{w_1^2 + w_2^2}$, $l_0 = 1$ mm, $k_s = 10$ N/mm, and $mg = 1$ N. The initial state of the system is $\lambda = 0$ N, $w_1 = 0$ mm, $w_2 = 1.1$ mm. Our objective is to use the proposed qANM to track how the equilibrium position $(w_1, w_2)$ evolves with respect to $\lambda$.

The q-Jacobi method in \cref{sec: linear} and the VQLS formulation in \cref{sec: VQLS} are separately combined with qANM to solve the nonlinear problem in the simulator. For the qANM parameters, we set the order of the Taylor series in \cref{eq: taylor} to $N = 10$, and the accuracy parameter $\epsilon_d$ in \cref{eq: amax} to $10^{-3}$.  The number of shots $n_s$ is set to $5 \times 10^5$ and 2 qubits are used for both quantum linear solvers. As a fully classical baseline, we also solve the same ANM systems with a classical linear solver using the same ANM parameters. For comparison, we also apply the Newton-Raphson (NR) method, combined with q-Jacobi for solving the linear equations arising from linearization, to solve the same problem. A detailed formulation of linearization via the NR method is provided in \ref{sec: NR}, where the termination condition for each nonlinear step is set as the residual norm $\epsilon_r = \| \bm{R}(\bm{u}^{(k+1)}, \bar{\lambda}) \| < 10^{-4}$ for this example.

\begin{figure}[!hbt]
\centering
\includegraphics[width=0.9\textwidth]{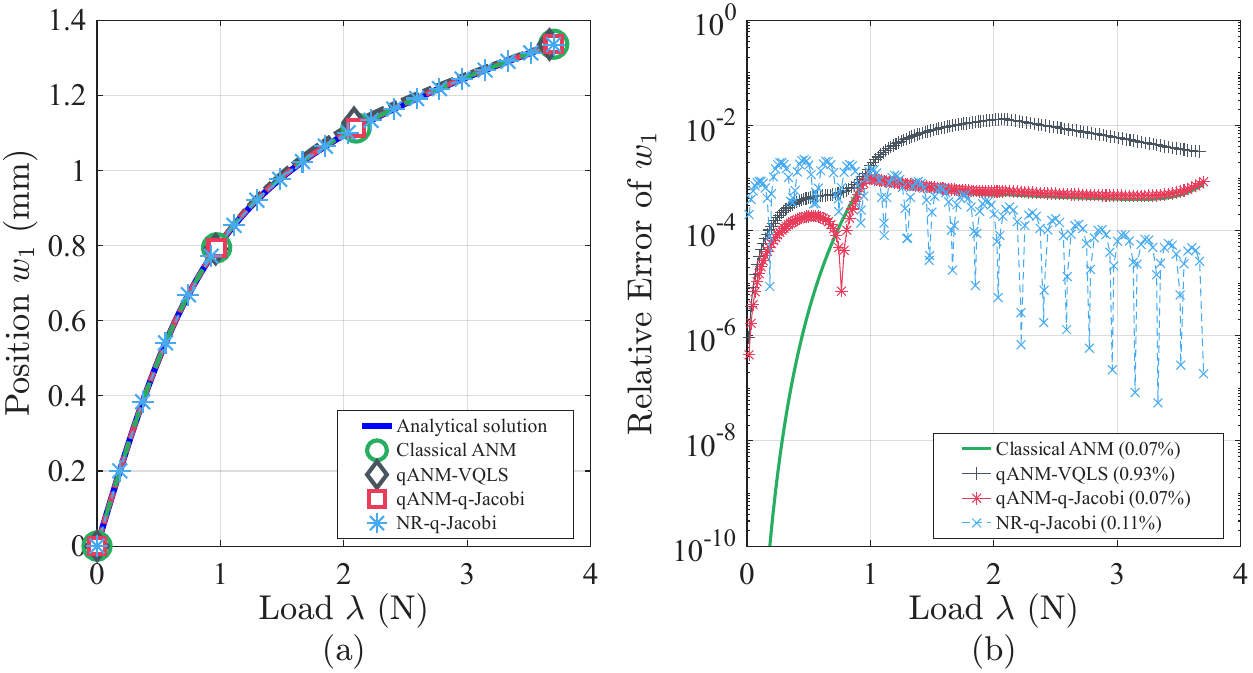}
\caption{(a) Solution paths $(w_1, \lambda)$ obtained by classical ANM, qANM combined with VQLS, qANM combined with q-Jacobi, and the NR method combined with q-Jacobi, together with the analytical solution. (b) Relative error of $w_1$ for each method compared with the analytical solution. The percentages in parentheses are the corresponding path errors. The markers on the numerical solution paths indicate the continuation step points.}
\label{fig: spring_two_qANM}
\end{figure}

\begin{figure}[!hbt]
\centering
\includegraphics[width=0.9\textwidth]{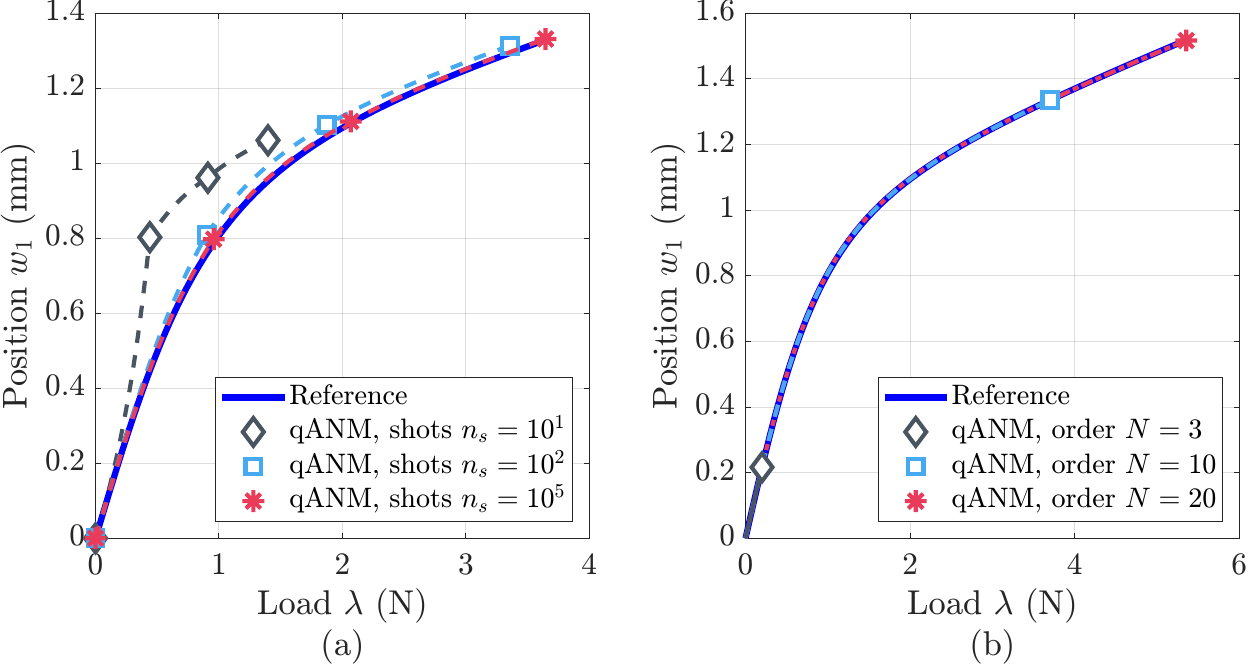}
\caption{(a) Solution paths obtained by qANM with different numbers of shots $n_s = 10^1$, $10^2$, and $10^5$, where the order $N$ is all set to 10. (b) Solution paths obtained by qANM with different Taylor series orders $N = 3$, $10$, and $20$, where the number of shots $n_s$ is all set to $5\times10^5$.}
\label{fig: spring_ns_N}
\end{figure}

\Cref{fig: spring_two_qANM}(a) presents the solution paths $(w_1, \lambda)$ obtained by classical ANM, qANM combined with VQLS, qANM combined with q-Jacobi, and the NR method combined with q-Jacobi, together with the analytical solution. Note that for each nonlinear step of qANM, we generate $100$ points on the solution path by uniformly sampling the path parameter $a \in [0,~ a_{max}]$, resulting in an approximately continuous solution path. For NR, after obtaining the discrete points (see \ref{sec: NR}) on the solution path, we generate new points corresponding to the same $\lambda$ of qANM using linear interpolation for comparison. \Cref{fig: spring_two_qANM}(b) shows the relative error of the obtained $w_1$ compared to the analytical solution. To evaluate the accuracy of the obtained path, we compute the path error defined as $\sqrt{\sum_{\lambda}{\left((w_1(\lambda)-\hat{w}_1(\lambda))\right)^2}/\sum_{\lambda}{(\hat{w}_1(\lambda))^2}} \times 100 \%$, where $\hat{w_1}(\lambda)$ represents a reference point on the solution path. All four methods achieve a path error of less than 1\%. Classical ANM gives a path error of $0.0700\%$, slightly lower than the $0.0736\%$ obtained by qANM combined with q-Jacobi; both values are displayed as $0.07\%$ in the legend. This small difference is expected because the classical linear systems are solved deterministically, whereas q-Jacobi uses estimates obtained with a finite number of shots that introduce small statistical errors into the Taylor coefficients. Notably, during the second continuation step, the error curves of classical ANM and qANM combined with q-Jacobi almost overlap. This close agreement indicates that q-Jacobi is sufficiently accurate in this interval to reproduce the path obtained by classical ANM. The path errors of qANM combined with VQLS and NR combined with q-Jacobi are $0.93\%$ and $0.11\%$, respectively. Both classical ANM and qANM require 3 nonlinear steps, whereas the NR method requires 20 steps. Thus, the large convergence region is inherited from the ANM formulation, while the quantum linear solvers replace the classical solution of the linear systems.

Moreover, we investigate the influence of the number of shots $n_s$ and the order of the Taylor series $N$ on the performance of qANM. Here, only the qANM combined with the q-Jacobi is presented for simplicity. \Cref{fig: spring_ns_N}(a) shows the solution paths obtained by qANM with $n_s = 10^1$, $10^2$, and $10^5$, where the order $N$ is all set to 10. The results show that the accuracy of qANM improves as $n_s$ increases. This aligns with the findings in \cref{sec: Evaluation}, where higher $n_s$ results in increased accuracy of the solution vector obtained by q-Jacobi. 

\Cref{fig: spring_ns_N}(b) displays the solution paths obtained by qANM with different Taylor series orders $N = 3$, $10$, and $20$, where the number of shots $n_s$ is all set to $5\times10^5$. All paths are calculated with 3 steps, but only the endpoint of the last step is displayed for clarity. The results indicate that increasing the order of the Taylor series can effectively enhance the convergence region of qANM.

In summary, the numerical results demonstrate that qANM is effective in solving nonlinear problems and exhibits a large convergence region, significantly reducing the number of nonlinear steps required to track a nonlinear solution path. Both VQLS and q-Jacobi can solve the linear systems generated by qANM, and both accurately reproduce the spring-mass response. An additional numerical test of qANM combined with q-Jacobi for the nonlinear deflection of an Euler--Bernoulli beam is provided in \cref{sec: Numerical}.

\subsection{Impact of solver inaccuracy on qANM}\label{sec: solver_inaccuracy}

\begingroup

The use of quantum linear solvers, which inherently produce approximate solutions, raises an important question regarding their integration with the qANM framework. The errors in these solutions, arising from sources such as algorithmic approximations and shot noise, can affect the performance of the method. Ideally, one would derive an analytical expression relating the accuracy of the nonlinear solution path to the quantum resources for a given problem dimension $D$. However, such a derivation is challenging for two primary reasons. First, the errors in the Taylor series coefficients are highly coupled. As shown in \cref{eq: linear_sys}, the right-hand side vector $\bm{F}_{\text{nl}}(\bm{u}_1, \dots, \bm{u}_{p-1})$ for order $p \ge 2$ depends on all previously computed solutions, i.e., $\bm{u}_1, \dots, \bm{u}_{p-1}$, causing errors to propagate and accumulate through the higher-order terms. Second, the automatic step-size mechanism, governed by \cref{eq: amax}, is sensitive to inaccuracies in the computed norms of $\bm{u}_1$ and $\bm{u}_N$, making the evolution of the step size difficult to predict analytically. We therefore turn to a numerical investigation to illustrate the impact of solver error on qANM.

We consider the buckling of an Euler-Bernoulli beam with strong geometric nonlinearity. The beam is modeled using Euler-Bernoulli kinematics and the von K\'{a}rm\'{a}n strain~\cite{reddy2014introduction}. The governing equations are
\begin{align*}
& \epsilon_{xx}=\frac{\partial u}{\partial x}+\frac{1}{2}\left(\frac{\partial w}{\partial x}\right)^2-z\frac{\partial^2 w}{\partial x^2},\\
& \sigma_{xx}=E\epsilon_{xx},\\
& \int_V \delta\epsilon_{xx}\sigma_{xx}\,dV-\delta W_{\mathrm{ext}}=0, 
\end{align*}
where $x$ is the coordinate along the beam axis, $z$ is the coordinate through the thickness, $u$ and $w$ are the displacements in the $x$- and $z$-directions, respectively, $\epsilon_{xx}$ is the von K\'{a}rm\'{a}n strain, $\sigma_{xx}$ is the stress, $V$ is the beam volume, and $\delta W_{\mathrm{ext}}$ is the external virtual work determined by the applied loading. After finite element discretization, these equations lead to a nonlinear algebraic system of the form \cref{eq: Nonlinear}. The corresponding analytical ANM coefficient construction for the Euler-Bernoulli beam discretization is provided in \cref{sec: EB_ANM_coefficients}. The same governing equations are used for the supplementary beam deflection problem in \cref{sec: Numerical}, with the external virtual work adapted to the uniform pressure applied there.

The boundary conditions, loading, and beam parameters are shown in \cref{fig: beam_buckling}. The beam is simply supported at both ends and subjected to compressive loads $\lambda f_0$. A small lateral load $\lambda f_0/10^7$ is applied at the midpoint to trigger buckling. The beam has length $L=30$ mm, width $B=1$ mm, height $H=1$ mm, and Young's modulus $E=3\times10^5$ MPa. The beam initially undergoes axial compression. As the compressive load approaches a critical value, a significant midpoint displacement occurs, marking the onset of buckling. The analytical critical compressive load is $\pi^2 EI/(\mu L)^2$ \cite{timoshenko2012theory}, where $I=BH^3/12$ and $\mu=1$ is the effective length factor for the simply supported boundary condition.

\begin{figure}[!htbp]
\centering
\includegraphics[width=8cm]{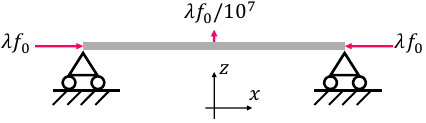}
\caption{Sketch of the Euler-Bernoulli beam, where the boundary condition triggers buckling with strong geometric nonlinearity.}
\label{fig: beam_buckling}
\end{figure}

The finite element discretization uses Lagrange interpolation for the axial displacement $u$ and Hermite cubic interpolation for the transverse displacement $w$ \cite{reddy2014introduction,choe2018efficient}. Owing to symmetry, only half of the beam is discretized, using 11 elements. For ANM, the Taylor series order is set to $N=16$, and the accuracy parameter is set to $\epsilon_d=10^{-8}$.

A separate classical comparison between ANM and the NR method for the same strongly nonlinear problem is provided in \cref{sec: adv_ANM}; it illustrates the path-following advantage inherited from the ANM formulation. We next return to the main focus of this section and investigate how shot noise in the linear system solutions affects qANM.

To this end, we simulate the effect of quantum solution error by focusing on shot noise, a dominant source of statistical error in quantum computing. For each linear system solve, we first compute the exact solution and then introduce noise that models the statistical process of quantum measurement. Specifically, for a solution vector $\bm{u}$, the noise added to each component is based on a binomial sampling process with $n_s$ shots, approximated by a normal distribution \cite{kuang2025quantum,li2017efficient}. For this example, we set $n_s=5\times 10^4$. The red curve in \cref{fig: Error_IR} shows the resulting solution path. It is clear that the accumulated error causes the path to diverge significantly from the reference solution. Furthermore, the automatic step-size control in \cref{eq: amax} fails, leading to an inefficient and unstable progression.

\begin{figure}[!tbp]
\centering
\includegraphics[width=0.6\textwidth]{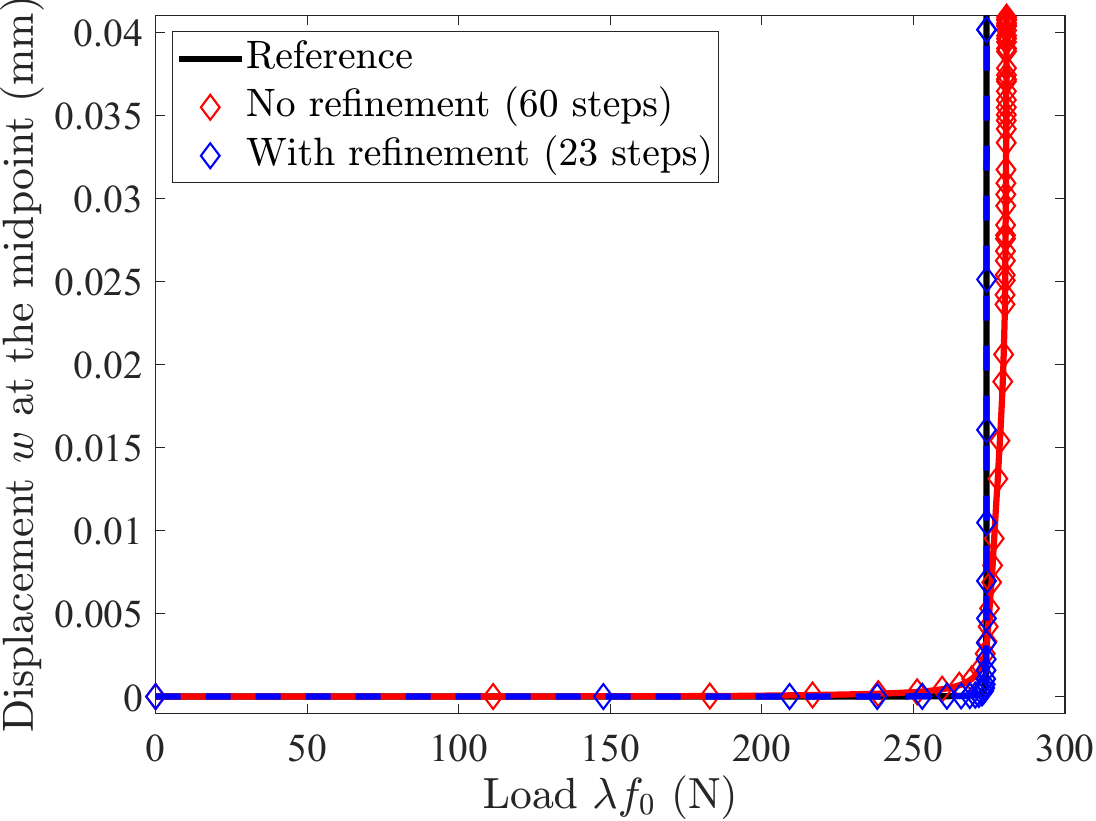}
\caption{Impact of solver error on qANM for the beam buckling problem. The red curve, with simulated shot noise ($n_s=5\times 10^4$), diverges from the reference solution. The blue curve, which uses iterative refinement to mitigate the noise, successfully tracks the reference solution and demonstrates effective automatic step-size control.}
\label{fig: Error_IR}
\end{figure}

To address this stability issue, one may use traditional error-correction techniques. A recent study has highlighted the potential of iterative refinement for enhancing the solutions from quantum linear solvers \cite{saito2023iterative}. This well-established method \cite{wilkinson1959rounding} proceeds as follows. Given an approximate solution $\tilde{\bm{u}}^{(i)}$ to the system $\bm{K}\bm{u}=\bm{F}$, one first computes the residual $\bm{r}^{(i)}=\bm{F}-\bm{K}\tilde{\bm{u}}^{(i)}$. Then, the correction equation $\bm{K}\Delta\bm{u}^{(i)}=\bm{r}^{(i)}$ is solved using the same noisy quantum solver and the same number of shots to find the correction term $\Delta\bm{u}^{(i)}$. The solution is then updated via $\tilde{\bm{u}}^{(i+1)}=\tilde{\bm{u}}^{(i)}+\Delta\bm{u}^{(i)}$. This process is repeated until the norm of the residual falls below a prescribed tolerance. Thus, the refinement uses the same noisy solver configuration and the same shot count, without increasing the assumed quantum resources per linear solve.

The blue curve in \cref{fig: Error_IR} demonstrates the effect of incorporating iterative refinement into the qANM procedure. Although this approach requires solving additional linear systems, it successfully stabilizes the algorithm. The resulting solution path agrees well with the reference, and the step-size control mechanism functions as intended, automatically adapting the step size to be larger in regions of low nonlinearity and smaller where the nonlinearity is strong. These results suggest that while the accuracy of quantum linear solvers is a critical factor for the stability of qANM, established techniques like iterative refinement can effectively mitigate these errors. The integration of such error mitigation strategies with quantum nonlinear solvers represents a promising direction for future research.
\endgroup

\section{Experiment on a superconducting quantum processor}\label{sec: Experiment}

In this section, we present a proof-of-principle demonstration of qANM executed on a superconducting quantum processor, to solve the spring-mass problem in \cref{sec: spring}. It is important to note that current quantum hardware is in the NISQ era \cite{preskill2018quantum}, meaning the accuracy of quantum computing is influenced by hardware noise. This hardware noise can arise from imperfections in gate operations and environmental disturbances, making it challenging to achieve high-precision results. The impact of the hardware noise on the performance of qANM will also be evaluated in the following.

\begin{figure}[!tb]
\centering
\includegraphics[width=0.6\textwidth]{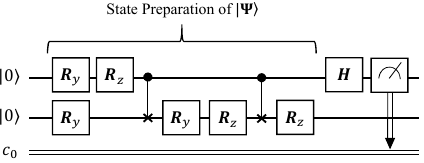}
\caption{Quantum circuit for solving the spring-mass problem using qANM, executed on the Quafu quantum processor.}
\label{fig: real_circuit}
\end{figure}

The experiments are conducted on the ScQ-P21 superconducting quantum processor from the Quafu Quantum Cloud Computing Cluster \cite{Quafu}, which comprises 21 qubits. Only two qubits are used in this work, where the single-qubit gate fidelity is 99.9\% and the two-qubit gate fidelity is 99.5\%.
For qANM, the order of the Taylor series expansion is set to $N = 4$, and the accuracy parameter is $\epsilon_d = 10^{-2}$. To implement the quantum linear solver q-Jacobi, the quantum state $\ket{\bm{\Psi}}$ is prepared using the \texttt{prepare\_state} method of \texttt{QuantumCircuit} in Qiskit~\cite{shende2004synthesis}. The resulting quantum circuit for solving the spring-mass problem is shown in \cref{fig: real_circuit}.  The q-Jacobi method is executed with $n_s = 5\times10^4$ shots, and the convergence criterion is set to $Tol < \epsilon_J = 10^{-3}$ or a maximum of 20 iterations.
For each inner product evaluated by q-Jacobi, the first part of the circuit in \cref{fig: real_circuit} generates the state $\ket{\bm{\Psi}}$ defined in \cref{eq: Phi_state}. Its amplitudes encode the normalized matrix row $\tilde{\bm{m}}_i$ and the current normalized Jacobi vector $\tilde{\bm{u}}^{(k)}$. Following Shende et al.~\cite{shende2004synthesis}, the target amplitude vector is divided into blocks containing two amplitudes. For the $c$-th block $\ket{\psi_c}$, described by the magnitude $r_c$, global phase $t_c$, polar angle $\theta_c$, and relative phase $\phi_c$, the corresponding rotations satisfy
\begin{equation*}
\bm{R}_y(-\theta_c)\bm{R}_z(-\phi_c)\ket{\psi_c}
=r_c e^{\mathrm{i}t_c}\ket{0}.
\end{equation*}
Here, the $\bm{R}_z$ rotation adjusts the relative phase, while the $\bm{R}_y$ rotation transfers the amplitude to $\ket{0}$. Multiplexed forms of these rotations act on all blocks and recursively disentangle one qubit at a time. Reversing and inverting this sequence therefore prepares $\ket{\bm{\Psi}}$ from $\ket{0}^{\otimes(n_q+1)}$, which explains the $\bm{R}_z$ and $\bm{R}_y$ gates in \cref{fig: real_circuit}. After the state has been prepared, the final $\bm{H}$ gate acts on the ancilla qubit as in the H-based inner-product algorithm of \cref{sec: linear}. Measuring the ancilla gives $P_0$, from which the required inner product is recovered using \cref{eq: inner_product_estimate}. This inner product gives the component $(\bm{M}\bm{u}^{(k)})_i$ through \cref{eq: component_recovery}, which is then used in the Jacobi update in \cref{eq: jacobi_iteration_simplified}.

The complete hardware experiment uses two ANM path steps and four Taylor orders, leading to 8 q-Jacobi linear solves and 63 Jacobi iterations. For the two-dimensional system, each Jacobi iteration submits one quantum circuit for each matrix row. The experiment therefore submits 126 quantum tasks, each using $n_s=5\times10^4$ shots, for a total of $6.3\times10^6$ shots. The total elapsed time recorded by the program is 3850.14~s, corresponding to averages of 30.56~s per submitted task, 61.11~s per Jacobi iteration, and 481.27~s per linear solve. Because only the total elapsed time was recorded, these values include circuit submission, compilation, possible queueing, execution, measurement, and result retrieval; the execution time on the processor cannot be separated. For comparison, the corresponding calculation performed locally on the classical computer takes approximately 0.007~s. Thus, the current hardware implementation is substantially slower than the classical computation. The measured runtime reflects the 126 separate cloud submissions and the repeated execution of $5\times10^4$ shots for each task, as well as the time required for gate operations, qubit reset, measurement, and result communication. For practical computational mechanics applications, a quantitative speedup over classical computing on current quantum hardware remains to be achieved \cite{preskill2018quantum}. Faster gate operations, qubit reset, and measurement may reduce these limitations in the future \cite{noiri2022fast,sunada2022fast}. 

\begin{figure}[!tb]
\centering
\includegraphics[width=0.65\textwidth]{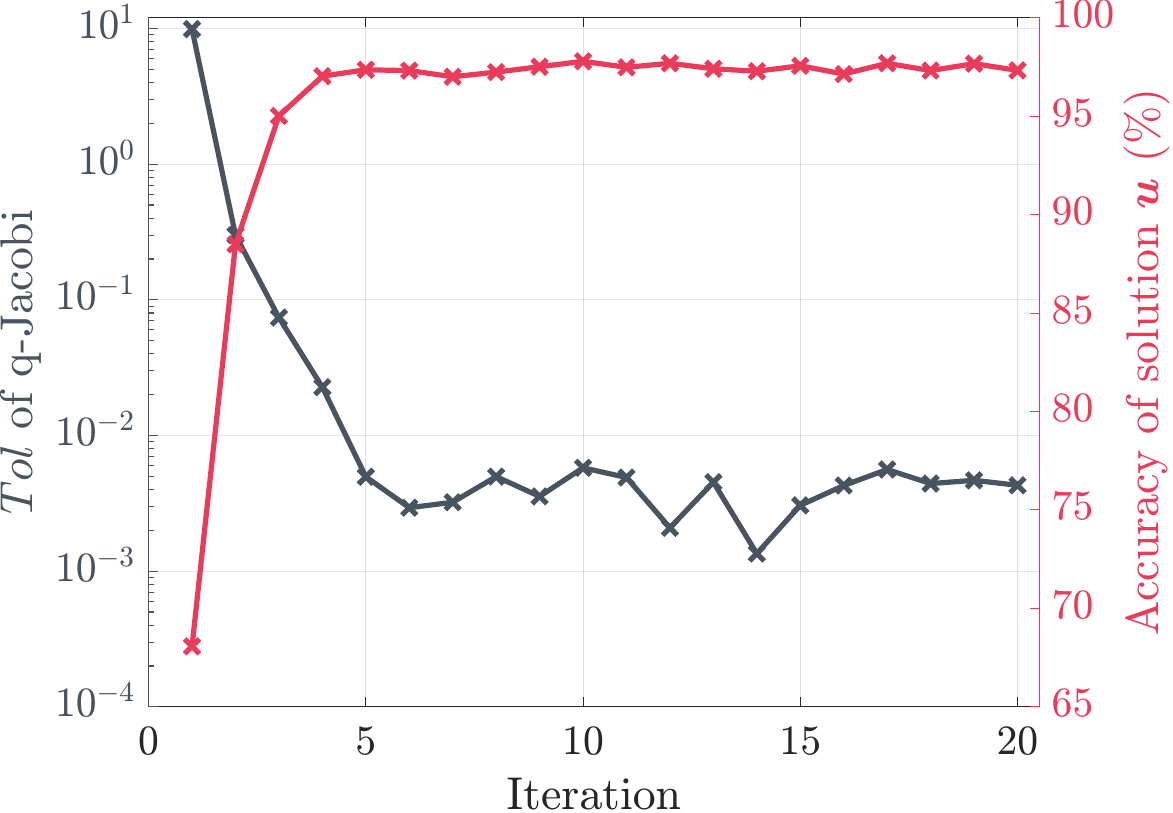}
\caption{Results of solving a linear equation using q-Jacobi on the Quafu quantum processor: evolution of $Tol$ and the accuracy of the solution vector obtained across iterations.}
\label{fig: real_iter}
\end{figure}

\begin{figure}[!tb]
\centering
\includegraphics[width=0.92\textwidth]{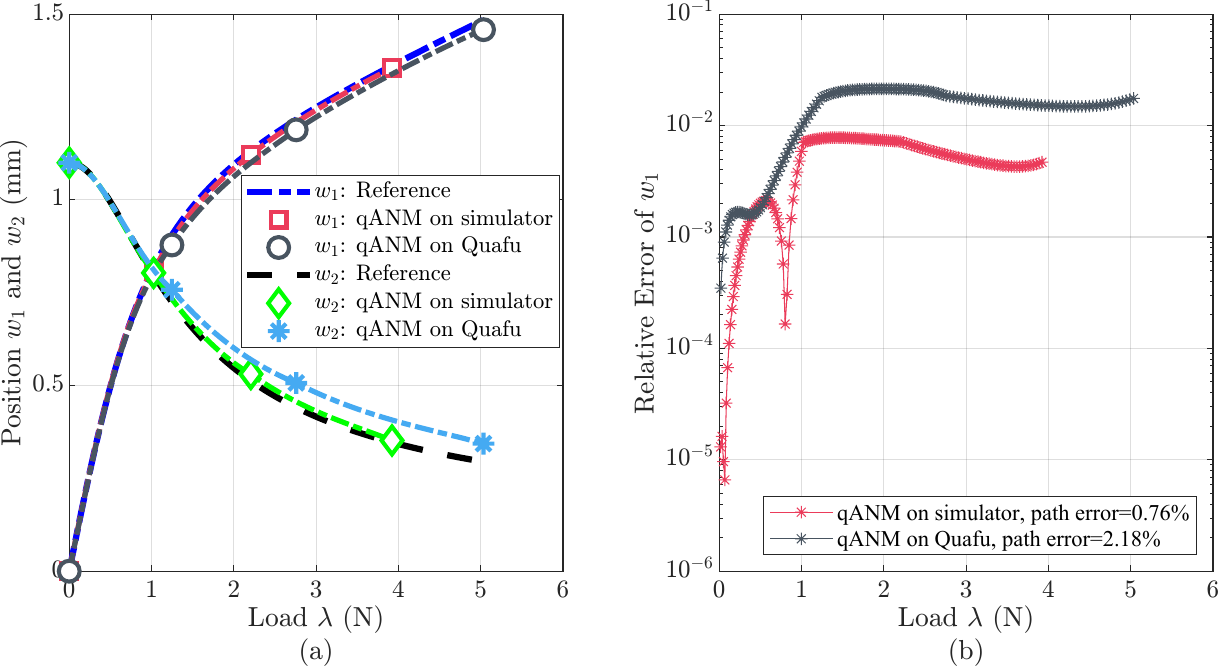}
\caption{Results for the spring-mass problem on the Quafu quantum processor: (a) Solution paths for $(w_1, \lambda)$ and $(w_2, \lambda)$, computed using qANM on the simulator, on the Quafu quantum processor, along with the reference analytical solution. (b) Relative error in $(w_1, \lambda)$ and overall path error for qANM on the simulator and Quafu quantum processor, compared to the reference solution.}
\label{fig: real_curves}
\end{figure}

\begin{figure}[!tbp]
\centering
\includegraphics[width=0.92\textwidth]{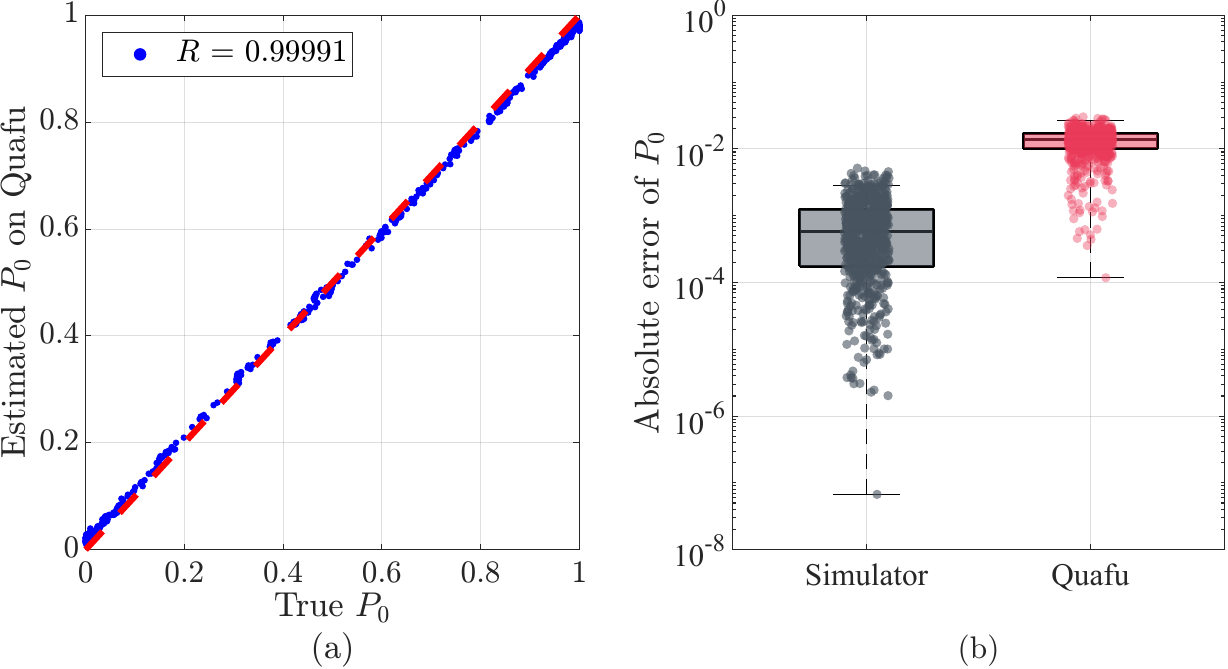}
\caption{Probability $P_0$ obtained from the Quafu quantum processor. (a) True $P_0$ versus the estimated $P_0$ on the Quafu quantum processor, along with the correlation coefficient $R$. (b) Comparison of the absolute error in the estimated $P_0$ obtained on the Quafu quantum processor and the Qiskit simulator, both with the same number of shots $n_s=5\times10^4$.}
\label{fig: real_error_P0}
\end{figure}

\cref{fig: real_iter} shows an example of solving the linear system $\bm{K} \bm{u} = \bm{F}$ on the quantum device, where $\bm{K}$ and $\bm{F}$ are obtained from the first-order in \cref{eq: linear_sys} at the first nonlinear step for the spring-mass problem. After 20 iterations, the q-Jacobi solver converges with an accuracy of about 97\%, which indicates the convergence of the q-Jacobi on the real quantum device. 
Furthermore,  \cref{fig: real_curves}(a) shows the solution paths of $(w_1, \lambda)$ and $(w_2, \lambda)$ for the spring-mass problem, computed using the reference analytical solution, qANM on the simulator, and qANM on the Quafu quantum processor. It should be emphasized that a key difference between the simulator and the Quafu quantum processor is that the former does not consider quantum hardware noise. Meanwhile, \cref{fig: real_curves}(b) presents the corresponding relative error in $(w_1, \lambda)$ as well as the overall path error, where the path error is defined the same as that in \cref{sec: spring}. Though not as accurate as the simulator (path error = 0.76\%), the results from the Quafu quantum processor achieve an accuracy of around 98\% (path error = 2.18\%). This demonstrates that qANM can be executed on a real quantum device and obtains a nonlinear solution path with a reasonable degree of accuracy in the NISQ era.

Finally, to evaluate the influence of quantum hardware noise, \cref{fig: real_error_P0}(a) shows all the $P_0$ (in \cref{eq: P0_derivation}) obtained from the Quafu quantum processor during the experiment, compared with the true $P_0$ exactly calculated using the equation on a classical computer. The correlation coefficient $R$ is 0.99991, which indicates the high accuracy of the Quafu quantum processor. In addition, \cref{fig: real_error_P0}(b) compares the absolute error in the estimated $P_0$ between the Quafu quantum processor and the Qiskit simulator, both using the same number of shots $n_s=5\times10^4$. On the simulator, since there is no quantum hardware noise, the averaged error is only around $1.2 \times 10^{-3}$, whereas on the Quafu quantum processor, it is significantly larger, approximately $1.3 \times 10^{-2}$. This increased error on the real device affects the accuracy of the q-Jacobi solver, which in turn impacts the overall accuracy of qANM. We believe that quantum error mitigation techniques, such as zero-noise extrapolation \cite{larose2022mitiq,temme2017error,li2017efficient,kandala2019error,kuang2025quantum}, could help improve the accuracy of qANM on real quantum devices in future research.

It should be noted that while the use of two qubits represents a small scale, solving nonlinear problems on real superconducting hardware remains a significant challenge due to noise and error accumulation. Existing demonstrations of nonlinear computations on quantum hardware remain limited to small problem sizes and specific formulations \cite{lubasch2020variational,umer2025probing,sarma2024quantum,chagneau2023quantum}. In this context, our experiment serves as a non-trivial proof-of-principle. The achievement of 98\% accuracy in tracking the solution path demonstrates the feasibility of the proposed algorithm in a realistic noisy environment. This experiment should be viewed as a small-scale proof-of-principle demonstration. It indicates that the proposed workflow can be executed on noisy superconducting hardware and can still recover the nonlinear solution path with reasonable accuracy.

Quantum computing is still in its early stage, and current hardware remains noisy and small in scale. Indeed, achieving efficient solutions of practical problems and demonstrating quantum advantage on real quantum hardware remains a recognized open challenge for the whole quantum computing community~\cite{preskill2018quantum}. The central value of this work is to demonstrate, on a real superconducting quantum processor, the complete nonlinear path-following solving process. In this process, a nonlinear equilibrium path of a mechanics problem is traced while the underlying linear systems are solved on actual quantum hardware. Although the demonstration is small in scale and no computational advantage is claimed, we believe that the proposed qANM framework and its analysis offer a useful reference for the future integration of quantum computing into nonlinear computational mechanics.

\section{Conclusion}\label{sec: Conclusion}

In this work, we presented the quantum asymptotic numerical method (qANM) for nonlinear path-following and demonstrated its implementation on a real superconducting quantum processor. Rooted in the principles of high-order perturbation theory, this approach uses Taylor series expansions to transform complex nonlinear behaviors into structured sequences of linear equations. To support the hardware implementation of qANM, we developed the quantum-enhanced Jacobi method (q-Jacobi), which uses a quantum subroutine to estimate the inner products required by the Jacobi iteration and serves as the quantum linear solver in the superconducting quantum processor experiment. Numerical tests confirmed that the high-order nature of the method leads to robust convergence, capturing the nonlinear solution paths effectively. A highlight of this work is the proof-of-principle experimental validation conducted on the superconducting quantum processor. While the scale is limited to two qubits, this implementation represents a non-trivial step considering the current hardware level. Despite the challenges posed by hardware noise in the current NISQ era, this work obtained a nonlinear solution path with 98\% accuracy. This result provides evidence that the qANM continuation procedure can be executed on noisy quantum hardware and can recover the nonlinear solution path with reasonable accuracy.

It is important to clarify the scope of the present contribution. This work should be viewed as a proof-of-principle study of nonlinear path-following in which the required linear systems are solved using quantum linear solvers, rather than as a demonstration of end-to-end quantum advantage. A practical quantum workflow for computational mechanics must account not only for the core quantum algorithm, but also for matrix access, state preparation, circuit construction, measurement cost, and output extraction. For q-Jacobi, efficient quantum state preparation remains a practical challenge, and the present implementation does not establish a quantum advantage over classical linear solvers. The current qANM implementation also requires the complete solution vector at each order of the Taylor expansion, because these coefficients are used to assemble higher-order terms and continue the nonlinear solution procedure. We therefore explicitly read out and update the vector classically, accepting the associated output cost. This choice reflects the well-known readout bottleneck of quantum computing~\cite{aaronson2015read}. These considerations define the current scope of the method and provide a clearer basis for evaluating its future development.

Looking forward, several directions emerge from this work. First, more advanced quantum linear-system algorithms, such as the quantum singular value transformation (QSVT) \cite{gilyen2019quantum}, may be considered in the qANM framework when the required matrix access and encoding procedures are available. Second, since qANM still uses ANM as a classical strategy to transform nonlinear problems into linear systems, an important open question is whether quantum computing can address nonlinear problems more directly, potentially drawing on recent advances in quantum simulation \cite{lloyd2020quantum,jin2023time,sato2024hamiltonian,meng2024simulating}. Third, the hardware results suggest that error mitigation \cite{larose2022mitiq,temme2017error,li2017efficient,kandala2019error} or error correction \cite{sun2022optical} will be important for extending such workflows to larger systems. Finally, the present study motivates further exploration of quantum computing approaches for nonlinear problems in mechanics, including instability, hyperelasticity, and elastoplasticity \cite{potier2024asymptotic}. Overall, the intersection of quantum computing and computational mechanics remains an important emerging direction, and this work provides a modest proof-of-principle reference for future developments.

\section*{Data availability}

Data and code used in this work will be made available on reasonable request.

\section*{Acknowledgements}

This work has been supported by the Key R\&D Program of Ningxia (Grant Number 2025BEE02006) and National Natural Science Foundation of China (Grant Numbers 12432009, 12172262, 12202322, and 12404578) . We would like to sincerely thank the support from the Synergetic Extreme Condition User Facility (SECUF) and the Quafu Quantum Cloud Computing Cluster \break (https://quafu.baqis.ac.cn/). 

\appendix
\renewcommand\thesection{\Alph{section}}
\titleformat{\section}
  {\large\bf}
  {\appendixname~\thesection}
  {1em}
  {}

\section{Additional details of the asymptotic numerical method}\label{sec: ANM_details}

\begingroup

The main text retains the ANM equations needed to understand how qANM converts a nonlinear problem into a sequence of linear systems and advances along the solution path. This appendix provides the remaining details for readers who wish to follow the coefficient calculation more closely, including the nonlinear terms, the arclength conditions, and the recursive construction of the Taylor coefficients.

We begin with a precise definition of $\bm{F}_{\mathrm{nl}}^{(p)}$. Let $[a^p]\{\cdot\}$ denote the coefficient of $a^p$ in the enclosed series. Then
\begin{equation}\label{eq: Fnl_coefficient}
\bm{F}_{\mathrm{nl}}^{(p)}
=
\left[
a^p
\right]
\left\{
\bm{R}
\left(
\bm{u}_0+\sum_{i=1}^{p-1}a^i\bm{u}_i,
\lambda_0+\sum_{i=1}^{p-1}a^i\lambda_i
\right)
-
\sum_{i=1}^{p-1}a^i
\left(
\bm{K}\bm{u}_i-\lambda_i\bm{F}
\right)
\right\},
\qquad 2\leq p\leq N.
\end{equation}
The coefficient extraction in \cref{eq: Fnl_coefficient} is performed analytically by matching equal powers of $a$; it does not use numerical differentiation or curve fitting. For two series $x(a)=\sum_{i\geq 0}a^i x_i$ and $y(a)=\sum_{j\geq 0}a^j y_j$, the coefficient of their product is
\begin{equation}\label{eq: analytical_product_rule}
[a^p]\{x(a)y(a)\}=\sum_{i=0}^{p}x_i y_{p-i}.
\end{equation}
The same rule applies componentwise to vector-valued multilinear terms. In particular, if $\Delta\bm{u}(a)=\sum_{i\geq 1}a^i\bm{u}_i$, then a bilinear term $\mathcal{Q}$ and a trilinear term $\mathcal{C}$ give
\begin{subequations}\label{eq: analytical_multilinear_rule}
\begin{align}
[a^p]\{\mathcal{Q}(\Delta\bm{u},\Delta\bm{u})\}
&=\sum_{i=1}^{p-1}\mathcal{Q}(\bm{u}_i,\bm{u}_{p-i}),\\
[a^p]\{\mathcal{C}(\Delta\bm{u},\Delta\bm{u},\Delta\bm{u})\}
&=\sum_{i=1}^{p-2}\sum_{j=1}^{p-i-1}
\mathcal{C}(\bm{u}_i,\bm{u}_j,\bm{u}_{p-i-j}).
\end{align}
\end{subequations}
Thus, coefficient extraction reduces to finite algebraic sums involving coefficients already known from lower orders. Since the load parameter enters linearly in the problems considered here, it produces only the term $-\lambda_p\bm{F}$ and no higher-order products. Once $\bm{u}_1,\ldots,\bm{u}_{p-1}$ are known, $\bm{F}_{\mathrm{nl}}^{(p)}$ is known and the $p$-th linear system in \cref{eq: linear_sys} can be solved. The explicit form for the Euler--Bernoulli beam discretization is derived in \cref{sec: EB_ANM_coefficients}.

At order $p$, \cref{eq: linear_sys} contains the vector $\bm{u}_p$ and the scalar $\lambda_p$, so one additional scalar equation is needed. It is supplied by the arclength definition in \cref{eq: arclength_definition}. Substituting \cref{eq: taylor} into this definition gives
\begin{equation}\label{eq: arclength_series}
a=
\sum_{p=1}^{N}a^p\left(\bm{u}_p^\top\bm{u}_1+\lambda_p\lambda_1\right).
\end{equation}
Equating the coefficients of equal powers of $a$ yields
\begin{equation}\label{eq: arclength_coefficients}
\bm{u}_1^\top\bm{u}_1+\lambda_1^2=1,
\qquad
\bm{u}_p^\top\bm{u}_1+\lambda_p\lambda_1=0,
\quad 2\leq p\leq N.
\end{equation}
These equations close the recursive construction of the Taylor coefficients.

For the first-order problem, let $\bar{\bm{u}}$ be the solution of
\begin{equation}\label{eq: ubar}
\bm{K}\bar{\bm{u}}=\bm{F}.
\end{equation}
Then $\bm{u}_1=\lambda_1\bar{\bm{u}}$. Using the first condition in \cref{eq: arclength_coefficients}, we obtain
\begin{equation}\label{eq: lambda1}
\lambda_1
=
\frac{\sigma}{\sqrt{1+\|\bar{\bm{u}}\|^2}},
\qquad
\bm{u}_1
=
\lambda_1\bar{\bm{u}},
\end{equation}
where $\sigma=\pm 1$ selects the orientation of the path. In practice, $\sigma$ is chosen according to the desired loading direction, for instance $\sigma=1$ for advancing along the positive load direction.

For each higher order $p\geq 2$, the nonlinear right-hand side $\bm{F}_{\mathrm{nl}}^{(p)}$ is already known from the previously computed coefficients. We first solve the auxiliary linear system
\begin{equation}\label{eq: uhat_p}
\bm{K}\hat{\bm{u}}_p
=
-\bm{F}_{\mathrm{nl}}^{(p)}.
\end{equation}
The general solution of the $p$-th equation in \cref{eq: linear_sys} can then be written as
\begin{equation}\label{eq: up_decomposition}
\bm{u}_p
=
\hat{\bm{u}}_p+\lambda_p\bar{\bm{u}}.
\end{equation}
Substituting this expression into the second condition in \cref{eq: arclength_coefficients} gives
\begin{equation}\label{eq: lambdap_derivation}
\left(\hat{\bm{u}}_p+\lambda_p\bar{\bm{u}}\right)^\top\bm{u}_1+\lambda_p\lambda_1=0,
\end{equation}
and therefore
\begin{equation}\label{eq: lambdap}
\lambda_p
=
-\frac{\hat{\bm{u}}_p^\top\bm{u}_1}
{\bar{\bm{u}}^\top\bm{u}_1+\lambda_1}.
\end{equation}
Since $\bm{u}_1=\lambda_1\bar{\bm{u}}$ and $\lambda_1^2(1+\|\bar{\bm{u}}\|^2)=1$, the denominator satisfies $\bar{\bm{u}}^\top\bm{u}_1+\lambda_1=1/\lambda_1$. Thus, the above expression can be simplified as
\begin{equation}\label{eq: lambdap_simplified}
\lambda_p
=
-\lambda_1\hat{\bm{u}}_p^\top\bm{u}_1,
\qquad
\bm{u}_p
=
\hat{\bm{u}}_p+\lambda_p\bar{\bm{u}}
=
\hat{\bm{u}}_p+\frac{\lambda_p}{\lambda_1}\bm{u}_1,
\quad 2\leq p\leq N.
\end{equation}
\cref{eq: ubar,eq: lambda1,eq: uhat_p,eq: lambdap_simplified} show explicitly how every coefficient $\bm{u}_p$ and $\lambda_p$ is obtained. The first-order coefficient determines the tangent direction of the solution branch, while the higher-order coefficients progressively correct the curvature of the branch. This recursive construction is the essential mechanism by which ANM replaces a nonlinear problem by a sequence of linear problems.

Once all coefficients are known, the local solution branch is represented by the polynomial map
\begin{equation}\label{eq: local_solution_path}
\bm{u}(a)=\bm{u}_0+\sum_{p=1}^{N}a^p\bm{u}_p,
\qquad
\lambda(a)=\lambda_0+\sum_{p=1}^{N}a^p\lambda_p.
\end{equation}
The scalar $a$ is therefore not an additional unknown solved from another nonlinear equation. It is the local coordinate along the equilibrium path. After the Taylor coefficients have been computed, any admissible value of $a$ gives a point on the continuous approximation of the solution branch.

\endgroup

\section{Analytical ANM coefficients for the Euler--Bernoulli beam}\label{sec: EB_ANM_coefficients}

\begingroup

This appendix derives the analytical form of $\bm{F}_{\mathrm{nl}}^{(p)}$ for the Euler--Bernoulli beam problems considered in this work. The derivation begins with the weak form and ends with the finite sums used at an arbitrary Taylor order $p$.

\subsection{Finite element form of the beam residual}
For an Euler--Bernoulli beam with the von K\'{a}rm\'{a}n strain, the axial strain is written as
\begin{equation}\label{eq: EB_strain_split}
\epsilon_{xx}=\epsilon_0-z\kappa,
\qquad
\epsilon_0=u_{,x}+\frac{1}{2}w_{,x}^2,
\qquad
\kappa=w_{,xx}.
\end{equation}
For a symmetric cross-section, integration over the cross-section gives the internal virtual work
\begin{equation}\label{eq: EB_virtual_work}
\delta W_{\mathrm{int}}
=\int_0^L
\left(EA\,\delta\epsilon_0\,\epsilon_0
+EI\,\delta\kappa\,\kappa\right)\,dx,
\end{equation}
where $A$ and $I$ are the cross-sectional area and second moment of area, respectively.

For a two-node beam element, let
\begin{equation}\label{eq: EB_element_vector}
\bm{u}^{e}
=\begin{bmatrix}
 u_1 & w_1 & \theta_1 & u_2 & w_2 & \theta_2
\end{bmatrix}^{\top},
\qquad
\theta_i=w_{,x}(x_i).
\end{equation}
The axial displacement is interpolated by linear Lagrange functions and the transverse displacement by cubic Hermite functions. Their derivatives can be written as
\begin{equation}\label{eq: EB_B_matrices}
u_{,x}=\bm{B}_u\bm{u}^{e},
\qquad
w_{,x}=\bm{B}_w\bm{u}^{e},
\qquad
w_{,xx}=\bm{B}_{\kappa}\bm{u}^{e}.
\end{equation}
Substitution into \cref{eq: EB_virtual_work} shows that the element internal-force vector contains linear, quadratic, and cubic terms:
\begin{equation}\label{eq: EB_element_residual}
\bm{r}_{\mathrm{int}}^{e}(\bm{u}^{e})
=\bm{K}_{\mathrm{L}}^{e}\bm{u}^{e}
+\mathcal{Q}^{e}(\bm{u}^{e},\bm{u}^{e})
+\mathcal{C}^{e}(\bm{u}^{e},\bm{u}^{e},\bm{u}^{e}),
\end{equation}
where the linear stiffness matrix and the two nonlinear operators are
\begin{subequations}\label{eq: EB_element_operators}
\begin{align}
\bm{K}_{\mathrm{L}}^{e}
&=\int_{L_e}
\left(EA\,\bm{B}_u^{\top}\bm{B}_u
+EI\,\bm{B}_{\kappa}^{\top}\bm{B}_{\kappa}\right)\,dx,\label{eq: EB_linear_stiffness}
\\
\mathcal{Q}^{e}(\bm{u}_i^{e},\bm{u}_j^{e})
&=\int_{L_e}EA\left[
\frac{1}{2}\bm{B}_u^{\top}
(\bm{B}_w\bm{u}_i^{e})(\bm{B}_w\bm{u}_j^{e})
+\bm{B}_w^{\top}
(\bm{B}_w\bm{u}_i^{e})(\bm{B}_u\bm{u}_j^{e})
\right]dx,
\\
\mathcal{C}^{e}(\bm{u}_i^{e},\bm{u}_j^{e},\bm{u}_k^{e})
&=\int_{L_e}\frac{EA}{2}\bm{B}_w^{\top}
(\bm{B}_w\bm{u}_i^{e})(\bm{B}_w\bm{u}_j^{e})
(\bm{B}_w\bm{u}_k^{e})\,dx.
\end{align}
\end{subequations}
Here, $\bm{u}_i^{e}$, $\bm{u}_j^{e}$, and $\bm{u}_k^{e}$ denote element-level Taylor coefficients. The bilinear operator $\mathcal{Q}^{e}$ need not be symmetric in its two arguments, whereas $\mathcal{C}^{e}$ is symmetric. After standard finite element assembly, the global residual takes the form
\begin{equation}\label{eq: EB_global_residual}
\bm{R}(\bm{u},\lambda)
=\bm{K}_{\mathrm{L}}\bm{u}
+\mathcal{Q}(\bm{u},\bm{u})
+\mathcal{C}(\bm{u},\bm{u},\bm{u})
-\lambda\bm{F}=\bm{0}.
\end{equation}

\subsection{Analytical extraction at Taylor order \texorpdfstring{$p$}{p}}
At the base point $\bm{u}_0$, write
\begin{equation}\label{eq: EB_increment_series}
\bm{u}(a)=\bm{u}_0+\Delta\bm{u}(a),
\qquad
\Delta\bm{u}(a)=\sum_{p=1}^{N}a^p\bm{u}_p.
\end{equation}
Here, $\bm{K}_{\mathrm{L}}$ is the linear stiffness matrix defined in \cref{eq: EB_linear_stiffness}, whereas $\bm{K}$ is the tangent matrix of the complete nonlinear residual evaluated at the current base point $\bm{u}_0$. Using the notation of \cref{eq: linear_sys}, its action on the $p$-th order coefficient is
\begin{equation}\label{eq: EB_tangent_operator}
\bm{K}\bm{u}_p
=\bm{K}_{\mathrm{L}}\bm{u}_p
+\mathcal{Q}(\bm{u}_0,\bm{u}_p)
+\mathcal{Q}(\bm{u}_p,\bm{u}_0)
+3\mathcal{C}(\bm{u}_0,\bm{u}_0,\bm{u}_p).
\end{equation}
Thus, $\bm{K}$ and $\bm{K}_{\mathrm{L}}$ are identical only when the nonlinear tangent contributions in \cref{eq: EB_tangent_operator} vanish, as at the undeformed initial base point.
Substituting \cref{eq: EB_increment_series} and the corresponding series for $\lambda(a)$ into \cref{eq: EB_global_residual}, and then collecting the coefficient of $a^p$, gives
\begin{equation}\label{eq: EB_Fnl_p}
\begin{split}
\bm{F}_{\mathrm{nl}}^{(p)}={}&
\sum_{i=1}^{p-1}
\mathcal{Q}(\bm{u}_i,\bm{u}_{p-i})
+3\sum_{i=1}^{p-1}
\mathcal{C}(\bm{u}_0,\bm{u}_i,\bm{u}_{p-i})\\
&+\sum_{i=1}^{p-2}\sum_{j=1}^{p-i-1}
\mathcal{C}(\bm{u}_i,\bm{u}_j,\bm{u}_{p-i-j}),
\qquad 2\leq p\leq N.
\end{split}
\end{equation}
The first sum in \cref{eq: EB_Fnl_p} is the quadratic contribution. The second sum contains the cubic terms with one base-point coefficient $\bm{u}_0$, and the double sum contains the cubic terms formed entirely from positive Taylor orders. For example,
\begin{subequations}\label{eq: EB_Fnl_examples}
\begin{align}
\bm{F}_{\mathrm{nl}}^{(2)}={}&
\mathcal{Q}(\bm{u}_1,\bm{u}_1)
+3\mathcal{C}(\bm{u}_0,\bm{u}_1,\bm{u}_1),\\
\bm{F}_{\mathrm{nl}}^{(3)}={}&
\mathcal{Q}(\bm{u}_1,\bm{u}_2)
+\mathcal{Q}(\bm{u}_2,\bm{u}_1)
+6\mathcal{C}(\bm{u}_0,\bm{u}_1,\bm{u}_2)
+\mathcal{C}(\bm{u}_1,\bm{u}_1,\bm{u}_1).
\end{align}
\end{subequations}
Every index on the right-hand side of \cref{eq: EB_Fnl_p} is smaller than $p$. Hence, the nonlinear vector is known when the $p$-th linear system is solved. The integrations defining $\bm{K}_{\mathrm{L}}$, $\mathcal{Q}$, and $\mathcal{C}$ are evaluated by standard Gaussian quadrature during finite element assembly. This numerical spatial integration is separate from the extraction of the Taylor order: the coefficient of $a^p$ is obtained analytically from the finite sums in \cref{eq: EB_Fnl_p}, without numerical differentiation or curve fitting.
\endgroup

\section{Linearization via the Newton-Raphson method}\label{sec: NR}

To compute the solution path $(\lambda, \bm{u})$ of the nonlinear system $\bm{R}(\bm{u}, \lambda) = 0$ in \cref{eq: Nonlinear}, the Newton-Raphson (NR) method proceeds iteratively by linearizing the system around an initial guess and updating the solution until convergence at each fixed value of $\lambda$. Starting from an initial guess $\bm{u}^{(0)}$ and a fixed parameter value $\lambda = \bar{\lambda}$, we aim to find the corresponding solution $\bar{\bm{u}}$.

At each iteration $k$, the NR method approximates the nonlinear system using a linear expansion around the current estimate $\bm{u}^{(k)}$:
\begin{equation}\label{eq: NR_linearization_corrected}
\bm{R}(\bm{u}^{(k)} + \Delta \bm{u}^{(k)}, \bar{\lambda}) \approx \bm{R}(\bm{u}^{(k)}, \bar{\lambda}) + \bm{K}^{(k)} \Delta \bm{u}^{(k)} = 0,
\end{equation}
where $\bm{K}^{(k)} = D_{\bm{u}} \bm{R}(\bm{u}^{(k)}, \bar{\lambda}) \in \mathbb{R}^{D \times D}$ is the Jacobian matrix of $\bm{R}$ with respect to $\bm{u}$, evaluated at $\bm{u}^{(k)}$ and $\lambda = \bar{\lambda}$.
We then formulate the linearized system to solve for $\Delta \bm{u}^{(k)}$:
\begin{equation}\label{eq: NR_linear_system_corrected}
\bm{K}^{(k)} \Delta \bm{u}^{(k)} = - \bm{R}(\bm{u}^{(k)}, \bar{\lambda}).
\end{equation}
Solving \cref{eq: NR_linear_system_corrected} for $\Delta \bm{u}^{(k)}$, we update the solution as follows:
\begin{equation}\label{eq: NR_update_u_corrected}
\bm{u}^{(k+1)} = \bm{u}^{(k)} + \Delta \bm{u}^{(k)}.
\end{equation}
This iterative process continues until convergence is achieved, typically when the residual norm $\epsilon_r=\| \bm{R}(\bm{u}^{(k+1)}, \bar{\lambda}) \|$ falls below a specified tolerance, resulting in the solution $\bar{\bm{u}}$ as $\bm{u}^{(k+1)}$.
By incrementing the target parameter $\bar{\lambda}$ and repeating the NR iterations for each value, we trace the solution path $(\lambda, \bm{u})$ as a series of discrete points. Unlike the ANM, which provides an analytical expression of the solution path through a Taylor series expansion (see \cref{sec: ANM}), the NR method yields discrete solutions and often requires more nonlinear steps due to its reliance on local linear approximations, which limit its convergence region.
In the spring-mass validation in \cref{sec: spring}, the NR method is combined with q-Jacobi so that the ANM and NR linearization strategies are compared using the same quantum linear solver. By contrast, the comparison between ANM and the NR method in \cref{sec: adv_ANM} uses classical linear solvers for both methods to isolate their path-following behavior. In both cases, the NR linear systems are obtained from \cref{eq: NR_linear_system_corrected}.

\section{Variational quantum linear solver}\label{sec: VQLS}

\begingroup

This appendix presents the variational quantum linear solver (VQLS) used in the numerical simulations. VQLS is included to illustrate that the linear systems generated by qANM can be solved using a quantum linear solver other than q-Jacobi. It is not used in the experiment on a superconducting quantum processor; the methodology in the main text and hardware implementation focus on q-Jacobi.

The variational quantum linear solver (VQLS) is a hybrid quantum-classical algorithm designed to find an approximate solution $\ket{\bm{u}}$ to the linear system in \cref{eq: KuF}, such that $\bm{K}\ket{\bm{u}} \propto \bm{F}$. The key components of VQLS are the ansatz, cost function, and classical optimizers, which will be introduced below.

First, VQLS represents the unknown vector $\bm{u}$ as a quantum state $\ket{\bm{u}(\bm{\theta})}$, generated through a parameterized quantum circuit known as an ansatz:
\begin{equation}\label{ansatz}
\ket{\bm{u}(\bm{\theta})} = \bm{V}(\bm{\theta}) |0\rangle^{\otimes n_q},
\end{equation}
where $n_q = \log_2 D$ is the number of qubits required, $\bm{V}(\bm{\theta})$ is the ansatz (a unitary operator) parameterized by a set of parameters $\bm{\theta}$, and $|0\rangle^{\otimes n_q}$ denotes the ground state of the qubits. An example of an ansatz is shown in \cref{fig: VQLS}(a), which is a hardware-efficient ansatz proposed in \cite{kandala2017hardware} and used in this work. In this example, the ansatz $\bm{V}(\bm{\theta})$ generates a 16-dimensional state vector ($D=16$) using 4 qubits.

Second, to quantify how closely $\ket{\bm{u}(\bm{\theta})}$ approximates the solution of \cref{eq: KuF}, a cost function is defined to measure the overlap between the left-hand side $\bm{K}\ket{\bm{u}(\bm{\theta})}$ and the right-hand side $\bm{F}$  \cite{bravo2023variational}:
\begin{equation}\label{cost}
C(\bm{\theta}) = \frac{\langle \bm{u}(\bm{\theta})| \bm{H}_L \ket{\bm{u}(\bm{\theta})}}{\langle \bm{u}(\bm{\theta})| \bm{K}^\dagger \bm{K} \ket{\bm{u}(\bm{\theta})}},
\end{equation}
where the effective Hamiltonian $\bm{H}_L$ is defined as:
\begin{equation}\label{HL}
\bm{H}_L = \bm{K}^\dagger \bm{U} \left( \mathbb{I} - \frac{1}{n_q} \sum_{j=1}^{n_q} |0_j\rangle \langle 0_j| \otimes \mathbb{I}_{\bar{j}} \right) \bm{U}^\dagger \bm{K},
\end{equation}
where $\bm{K}^\dagger$ represents the conjugate transpose of $\bm{K}$, and $\bm{U}$ is a unitary operator that prepares the quantum state corresponding to the normalized vector $\bm{F}$, such that $|\bm{F}\rangle = \bm{U} |0\rangle^{\otimes n_q}$.
The term $\mathbb{I}$ denotes the identity operator, and $|0_j\rangle \langle 0_j| \otimes \mathbb{I}_{\bar{j}}$ is a projection operator acting on all qubits except $j$. In addition, VQLS assumes that $\bm{K}$ can be efficiently decomposed into a linear combination of unitary matrices, such that $\bm{K} = \sum_{i=0}^{k} c_i \bm{G}_i$, where $\bm{G}_i$ are unitary matrices and $c_i$ are real coefficients. The quantum circuit for computing the cost function $C(\bm{\theta})$ is illustrated in \cref{fig: VQLS}(b), and the total number of qubits required is $n_q + 1$. 

Third, by minimizing $C(\bm{\theta})$ with respect to the parameters $\bm{\theta}$ using a classical optimizer, e.g., gradient descent \cite{bravo2023variational} or COBYLA \cite{powell1998direct}, VQLS converges to an optimal set of parameters that generates the approximate solution. Specifically, the algorithm begins with an initial guess for the parameters $\bm{\theta}$. On the quantum computer, the cost function $C(\bm{\theta})$ is computed using the circuit shown in \cref{fig: VQLS}(b). This cost value is then returned to the classical optimizer, which iteratively updates $\bm{\theta}$ to minimize $C(\bm{\theta})$. This process repeats until convergence criteria are met, resulting in an optimized $\ket{\bm{u}(\bm{\theta})}$ that approximates a normalized solution to \cref{eq: KuF}. To recover the obtained $\ket{\bm{u}(\bm{\theta})}$, i.e., a unit quantum state vector, to the original scale of the true solution $\bm{u}$, one can perform least-squares fitting to find a scale factor $s$ that minimizes the $l_2$-norm $|| \bm{F} - s \bm{K} \ket{\bm{u}(\bm{\theta})} ||$ \cite{chen2024enabling}. The scale factor can be easily computed as  $s=(\bm{F}^\top \bm{K}\ket{\bm{u}(\bm{\theta})}) / ((\bm{K}\ket{\bm{u}(\bm{\theta})})^\top (\bm{K}\ket{\bm{u}(\bm{\theta})}))$, then the final quantum solution is obtained as $\bm{u} = s \ket{\bm{u}(\bm{\theta})}$.

\begin{figure}[!tb]
\centering
\includegraphics[width=0.7\textwidth]{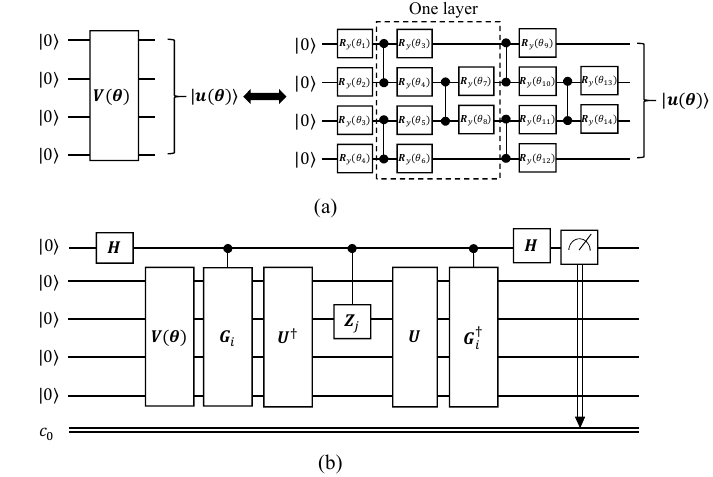}
\caption{(a) Illustration of the hardware-efficient ansatz \cite{kandala2017hardware}. The ansatz begins with single-qubit $\bm{R}_y$ rotations, followed by layers of entangling gates and parameterized rotations. Each $\bm{R}_y$ gate is associated with a parameter in $\bm{\theta}$, which is optimized during the algorithm. (b) Quantum circuit for computing the cost function $C(\bm{\theta})$ in VQLS.}
\label{fig: VQLS}
\end{figure}

Under suitable assumptions on matrix access, state preparation, and measurement, VQLS has been reported to exhibit favorable scaling with the system dimension, often expressed as $O(\mathrm{polylog}(D))$ for the quantum linear-solver subroutine~\cite{bravo2023variational}. When qANM is combined with VQLS, this subroutine is applied to the sequence of linear systems generated by the ANM expansion. Since $N$ linear systems are solved within one continuation step, as shown in \cref{eq: linear_sys}, the corresponding subroutine-level cost can be formally written as $O(N\mathrm{polylog}(D))$. In practical ANM computations, the Taylor series order $N$ is usually moderate, often chosen as $N\leq 20$~\cite{guillot2019generic}. However, this estimate should be interpreted carefully. It does not include the cost of constructing or decomposing the tangent matrix, preparing the right-hand sides, performing measurements, or extracting the solution information required by the ANM recursion. Therefore, the VQLS component should be viewed here as a possible quantum linear-solver module within qANM, rather than as a complete demonstration of end-to-end computational advantage.
\endgroup

\section{Numerical test: deflection of an Euler-Bernoulli beam}\label{sec: Numerical}

\begingroup

This appendix presents an additional numerical test of qANM for the geometrically nonlinear deflection of an Euler-Bernoulli beam \cite{reddy2014introduction}. The analytical ANM coefficient construction for this beam discretization is provided in \cref{sec: EB_ANM_coefficients}.

\begin{figure}[!hbtp] \centering \includegraphics[width=6cm]{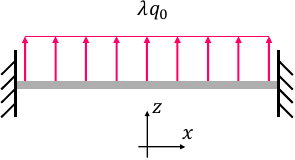} \caption{Sketch of the Euler-Bernoulli beam, where the upper surface is applied with uniform pressure.} \label{fig: beam} \end{figure}

\begin{figure}[!htp] 
\centering 
\includegraphics[width=0.5\textwidth]{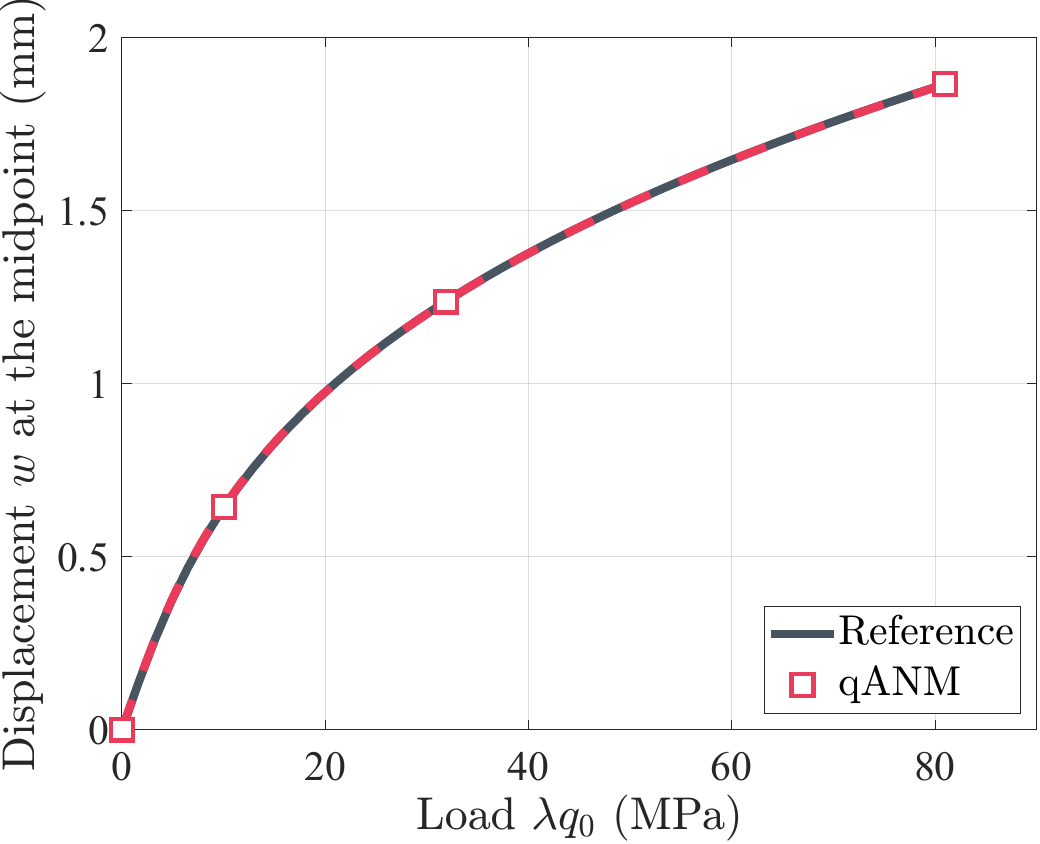} 
\caption{Displacement $w$ at the midpoint of the Euler-Bernoulli beam, comparing results obtained using the proposed qANM and the reference classical NR method.} 
\label{fig: w_beam} 
\end{figure}

\begin{figure}[!ht] 
\centering 
\includegraphics[width=0.7\textwidth]{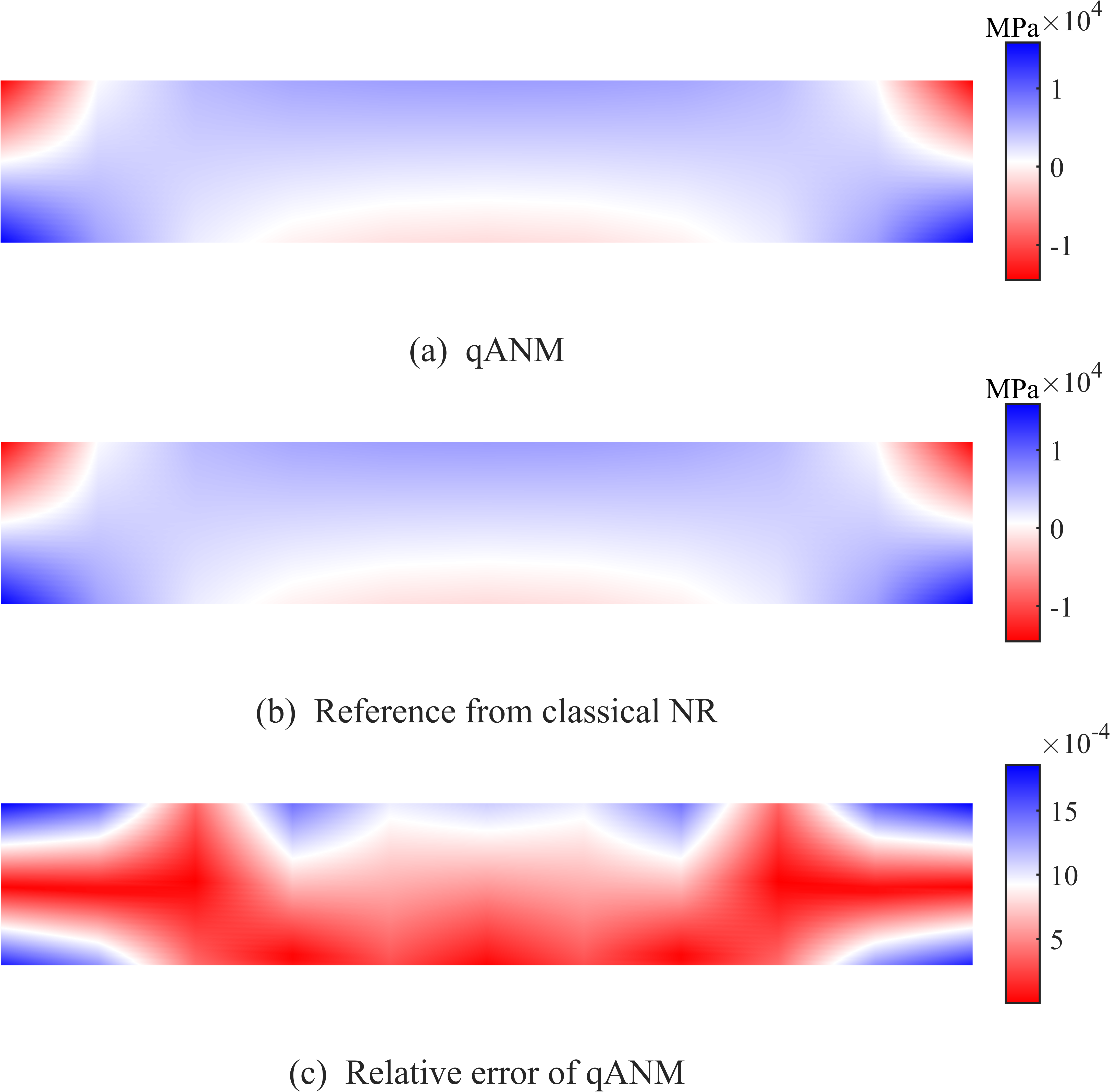} 
\caption{Stress distribution $\sigma_{xx}$ along the Euler-Bernoulli beam at the final solution point, comparing results from the qANM and the reference computed by the classical NR method. Note that the height of the displayed beam has been magnified by a factor of 5 for better visualization.} 
\label{fig: stress_beam} 
\end{figure}

The sketch of the Euler-Bernoulli beam is illustrated in \cref{fig: beam}, where both ends of the beam are fixed, and a uniform pressure of $\lambda q_0$, with $q_0=100$ MPa, is applied to the upper surface. The beam has a length of $L=30$ mm, a width of $B=1$ mm, and a height of $H=1$ mm. The material is assumed to be linear elastic with Young's modulus $E=3\times10^5$ MPa.

The governing Euler-Bernoulli beam equations are given in \cref{sec: solver_inaccuracy}. For the uniform pressure considered here, the external virtual work in that formulation is
\[
\delta W_{\mathrm{ext}}=\int_0^L \delta w\,\lambda q_0 B\,dx.
\]
After finite element discretization, the problem takes the nonlinear algebraic form given in \cref{eq: Nonlinear} and is solved using qANM.

The details of the numerical setup are as follows. We use the Lagrange interpolation function for discretizing $u$ and Hermite cubic interpolation functions for $w$ \cite{reddy2014introduction,choe2018efficient}. Due to the symmetry of the problem, only half of the beam is discretized, using 5 elements. Considering the boundary conditions, the resulting linear system has a dimension of $D = 13$. For qANM, the order of the Taylor series expansion is set to $N = 8$, and the accuracy parameter is $\epsilon_d = 10^{-5}$. For the quantum linear solver q-Jacobi, the number of shots is set to $n_s = 10^8$, with a termination condition of $Tol < \epsilon_J = 10^{-4}$ and 5 qubits are used for this problem.

\cref{fig: w_beam} shows the evolution of the displacement $w$ at the midpoint of the beam. The results from qANM are compared with a reference solution obtained using the classical NR method. After three nonlinear iterations, qANM yields a continuous solution path that closely matches the reference solution. Additionally, \cref{fig: stress_beam} presents the distribution of stress $\sigma_{xx}$ corresponding to the last solution point shown in \cref{fig: w_beam}, with a load of $\lambda q_0 = 80.96$ MPa. Note that the height of the displayed beam has been magnified by a factor of 5 for better visualization. The reference stress distribution is also obtained using the classical NR method. The results demonstrate that the stress field computed using qANM exhibits a maximum relative error of only around $2 \times 10^{-3}$, highlighting the effectiveness of the qANM in solving the nonlinear Euler-Bernoulli beam problem.

In summary, this supplementary example shows that qANM accurately solves the nonlinear Euler-Bernoulli beam deflection problem. It is included in the appendix as additional validation, while the Validation section in the main text focuses on the examples most directly connected to quantum linear-solver performance, solver inaccuracy, iterative refinement, and the experiment on quantum hardware.

\endgroup

\section{Classical comparison of ANM and the NR method for beam buckling}\label{sec: adv_ANM}

\begingroup

This appendix compares the classical ANM and Newton--Raphson (NR) methods for the Euler-Bernoulli beam buckling problem introduced in \cref{sec: solver_inaccuracy}. The analytical ANM coefficient construction for this beam discretization is provided in \cref{sec: EB_ANM_coefficients}. The purpose is to isolate the contribution of the high-order ANM path-following formulation. Since all linear systems considered in this appendix are solved classically, the comparison provides evidence for the robustness of the ANM formulation itself rather than validation of a quantum linear solver.

The beam configuration, loading, material parameters, finite element discretization, and ANM settings are given in \cref{sec: solver_inaccuracy}. For the NR method, the termination condition for each nonlinear step is set as the residual norm $\epsilon_r=\|\bm{R}(\bm{u}^{(k+1)},\bar{\lambda})\|<10^{-9}$; the linearization procedure is summarized in \cref{sec: NR}. The ANM calculations in this comparison use a classical linear solver rather than q-Jacobi because the very small lateral load $\lambda f_0/10^7$ requires greater accuracy in the solution vector than the current q-Jacobi implementation can provide. A quantum linear solver with higher accuracy would be required before qANM could be applied directly to this strongly nonlinear benchmark.

\begin{figure}[!tbp]
\centering
\includegraphics[width=0.55\textwidth]{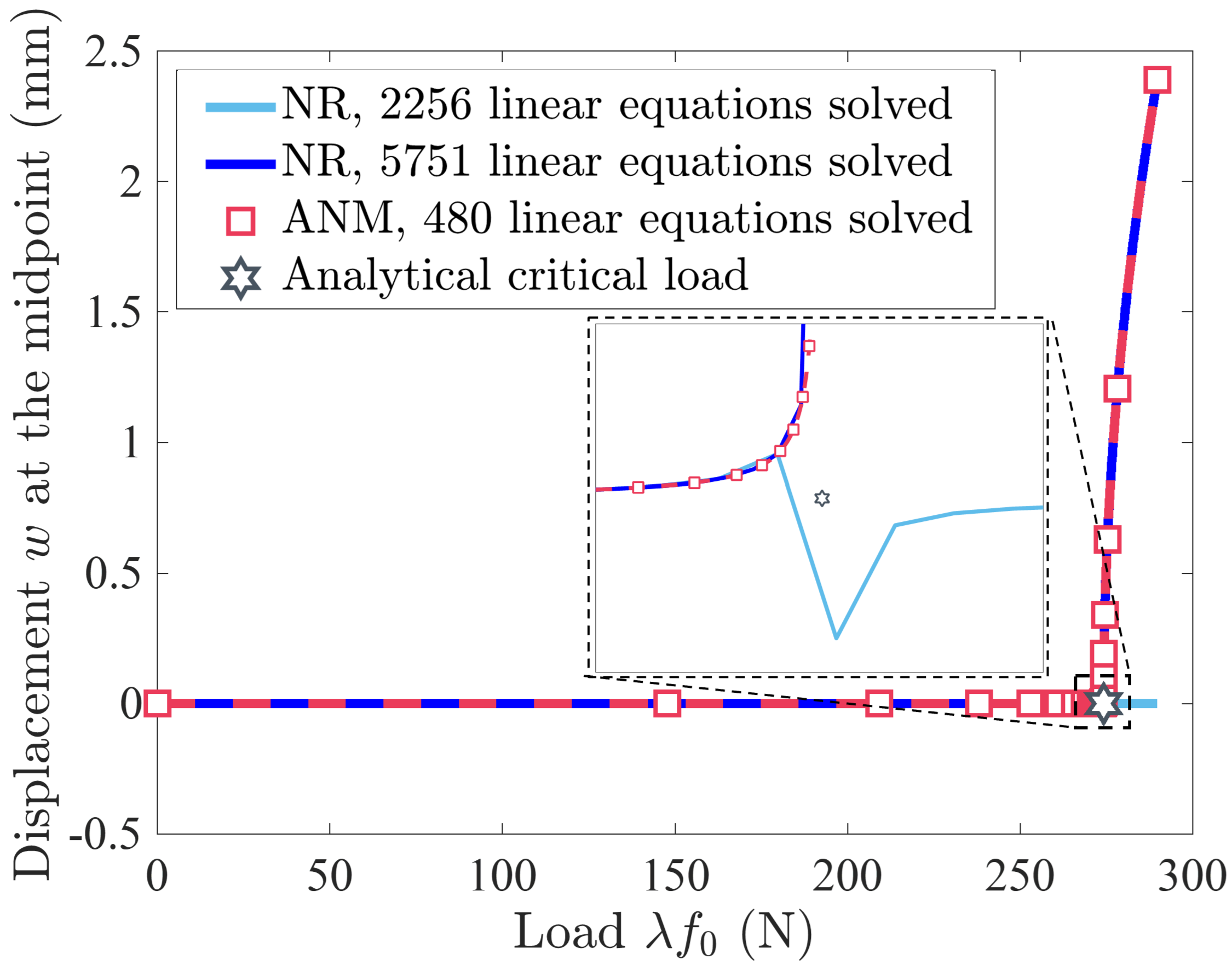}
\caption{Displacement $w$ at the midpoint of the Euler-Bernoulli beam, obtained using ANM and the NR method.}
\label{fig: Euler_dis}
\end{figure}

\Cref{fig: Euler_dis} shows the midpoint displacement $w$ calculated using both ANM and the NR method. With 30 nonlinear steps, ANM predicts the buckling phenomenon and requires the solution of 480 linear equations. In contrast, the NR method fails to predict buckling with 2000 nonlinear steps and 2256 linear equation solves. The NR method requires at least 4500 nonlinear steps and 5751 linear equation solves to capture the buckling phenomenon accurately. In this example, the ANM formulation therefore tracks the strongly nonlinear solution path with substantially fewer continuation steps than the NR method. Since qANM adopts this ANM formulation to construct the nonlinear solution path, this robustness is beneficial to the overall workflow. However, the result should be interpreted as evidence for the ANM path-following formulation rather than as a demonstration of quantum computational advantage.
\endgroup

\end{document}